\documentclass[traditabstract,proofs]{aa} 
\usepackage{graphicx}
\usepackage{hyperref}
\usepackage{latexsym}
\usepackage{color}
\usepackage{amsmath,amssymb}

\usepackage{natbib}
\usepackage{soul}
\usepackage{txfonts}
\usepackage{float}
\usepackage{longtable,lscape}
\newcommand{\degree}{$^{\circ}$}
\newcommand{\teff}{$T_{\rm eff}$}
\newcommand{\logg}{$\log g$}
\newcommand{\vsini}{$v\sin i$}

\newcommand{\rotfit}{{\sf ROTFIT}}

\newcommand{\COMPO}{{\sf COMPO2}}
\newcommand{\kms}{km\,s$^{-1}$}
 
\newcommand{\gaia}{{\it Gaia}}

\newcommand{\Lacc}{{$L_{\rm acc}$}}
\newcommand{\Macc}{{$\dot{M}_{\rm acc}$}}

\definecolor{blu}{rgb}{0,0,1}
\definecolor{mag}{rgb}{1,0,1}

\begin{document}

\title{PENELLOPE II. CVSO 104: a pre-main sequence close binary \\with an optical companion in Ori OB1\thanks{Based on data obtained within ESO programme 106.20Z8.}}

\author{A. Frasca\inst{\ref{OACT}} 
\and    H. M. J. Boffin\inst{\ref{instESO}}
\and    C. F. Manara\inst{\ref{instESO}}
\and    J. M. Alcal\'a\inst{\ref{instNA}}
\and    P. \'Abrah\'am\inst{\ref{instKO},\ref{instEL}} 
\and    E. Covino\inst{\ref{instNA}}
\and    M. Fang\inst{\ref{pmo}}
\and    M. Gangi\inst{\ref{instRO}}
\and    G. J. Herczeg\inst{\ref{KIAA}}
\and    \'A. K\'osp\'al\inst{\ref{instKO},\ref{instEL},\ref{instHD}} 
\and    L. Venuti\inst{\ref{instSETI}} 
\and    F.~M. Walter\inst{\ref{instSBU}}
\and    J. Alonso-Santiago\inst{\ref{OACT}}
\and    K. Grankin \inst{\ref{instCrAO}}
\and    M. Siwak\inst{\ref{instKO}} 
\and    E. Alecian\inst{\ref{instIPAG}}
\and    S. Cabrit\inst{\ref{ObsParis}} 
       } 

\institute{INAF -- Osservatorio Astrofisico di Catania, via S. Sofia 78, 95123 Catania, Italy \label{instCT}\\ \email{antonio.frasca@inaf.it}\label{OACT}
\and
European Southern Observatory, Karl-Schwarzschild-Strasse 2, 85748 Garching bei M\"unchen, Germany \label{instESO} 
\and
INAF -- Osservatorio Astronomico di Capodimonte, via Moiariello, 16, 80131 Napoli, Italy \label{instNA}
\and
Konkoly Observatory, Research Centre for Astronomy and Earth Sciences, E\"otv\"os Lor\'and Research Network (ELKH), Konkoly-Thege Mikl\'os \'ut 15-17, 
H-1121 Budapest, Hungary\label{instKO}
\and
ELTE E\"otv\"os Lor\'and University, Institute of Physics, P\'azm\'any P\'eter s\'et\'any 1/A, H-1117 Budapest, Hungary\label{instEL}
\and
Purple Mountain Observatory, Chinese Academy of Sciences, 10 Yuanhua Road, Nanjing 210023, China \label{pmo}
\and
INAF -- Osservatorio Astronomico di Roma, via di Frascati 33, 00078 Monte Porzio Catone, Italy\label{instRO}
\and
Kavli Institute for Astronomy and Astrophysics, Peking University, Yiheyuan 5, Haidian Qu, 100871 Beijing, China\label{KIAA}
\and
Max Planck Institute for Astronomy, K\"onigstuhl 17, D-69117 Heidelberg, Germany\label{instHD}
\and
SETI Institute, 189 Bernardo Ave, Suite 200, Mountain View, CA 94043, USA\label{instSETI}
\and
Stony Brook University, Stony Brook, NY 11794, USA
\label{instSBU}
\and
Crimean Astrophysical Observatory, 298409 Nauchny, Crimea \label{instCrAO}
\and 
Univ. Grenoble Alpes, CNRS, IPAG, 38000 Grenoble, France
\label{instIPAG}
\and
Observatoire de Paris, PSL University, Sorbonne Universit\'e, CNRS, LERMA, F-75014 Paris
\label{ObsParis}
}

\date{Received  / Accepted }

\abstract{We present results of our study of the close pre-main sequence spectroscopic binary CVSO\,104 in Ori OB1, based on data obtained within the PENELLOPE 
legacy program. We derive, for the first time, the orbital elements of the system and the stellar parameters of the two components. The system is composed of 
two early M-type stars and has an orbital period of about 5 days and a mass ratio of 0.92, but contrarily to expectations does not appear to have a tertiary 
companion. Both components have been (quasi-)synchronized, but the orbit is still very eccentric. The spectral energy distribution clearly displays a significant 
infrared excess compatible with a circumbinary disk. The analysis of  \ion{He}{i} and Balmer line profiles, after the removal of the composite photospheric spectrum, 
reveals that both components are accreting at a similar level. We also observe excess emission in H$\alpha$ and H$\beta$, which appears redshifted or blueshifted 
by more than 100 \kms\ with respect to the mass center of the system depending on the orbital phase. This additional emission could be connected with accretion 
structures, such as funnels of matter from the circumbinary disk.  
We also analyze the optical companion located at about 2$\farcs$4 from the spectroscopic binary. 
This companion, that we named CVSO~104\,B, turns out to be a background Sun-like star not physically associated with the PMS system and not belonging to Ori OB1.
} 

   \titlerunning{A close pre-main sequence binary in Ori OB1}
      \authorrunning{A. Frasca et al.}

\maketitle

\section{Introduction}
\label{Sec:intro}
The formation of stars and planets is strongly influenced by the conditions in their environment (circumstellar disks, jets and winds) during their early life.
The majority of stars are formed in binary and multiple systems, which, for sufficiently small separations, allows obtaining information on their components, 
such as dynamical masses.
It is therefore very important to detect and study young pre-main sequence (PMS) close binaries for deriving their orbital and stellar properties and to 
understand how the binarity affects the planet formation.
Moreover, well-characterized early-PMS binary systems are critical to constrain theoretical evolutionary tracks, which 
are used to derive fundamental stellar properties, but are often affected by offsets of up to 50\% in the predicted masses
compared to the dynamical mass estimates \citep[e.g.,][]{Covino2004,Hillenbrad2004,Stassun2014}. 
PMS binaries are also very important for studying the mass accretion process in a different field geometry from that of single stars.
Both observations and numerical simulations show that quasi-periodic bursts of accretion are expected in close binaries, with a different pattern for circular 
and eccentric systems  \citep[e.g.,][]{Munoz2016,Gillen2017}.
For eccentric systems, the accretion bursts, originating from the impact of nearly free-falling matter on the high atmospheric layers of each of the two components, 
are more frequently observed near the periastron passages \citep[e.g.,][]{Kospal2018,Tofflemire2019}. The simulations also show  gas streams from the circumbinary 
disk flowing to each component through the $L_2$ and $L_3$ Lagrangian points \citep[e.g.,][]{Val-Borro2011,Munoz2016}.

We present here the spectroscopic orbit of CVSO~104 (\object{Haro 5-64}), which is a classical T Tauri star located in the $\sim$ 5 Myr old Orion OB1b association. 
It was discovered as a star with H$\alpha$ emission in the region of the Horsehead Nebula (IC\,434) by \citet{Haro1953} with objective-prism observations. 
It was detected in the Kiso survey of emission-line objects by \citet{1989PASJ...41..155W} who confirmed a strong H$\alpha$ emitter.
\citet{Briceno2005} observed this star as part of the CIDA Variability Survey of Orion OB1 (CVSO) and reported it as a variable source with H$\alpha$ and 
\ion{Li}{i}\,$\lambda$6708 equivalent widths of $-62.90$\,\AA\ and 	0.30\,\AA, respectively.	

The spectra analyzed in the present paper have been obtained as part of the PENELLOPE program running at the Very Large Telescope (VLT). PENELLOPE is a ground-based 
follow-up large program \citep[see][hereafter Paper\,I]{Manara21}  of the ULLYSES HST program\footnote{\tt ullyses.stsci.edu} \citep{ullysesDR1}. 

The {\it Gaia}\,EDR3 release lists a parallax $\varpi  = 2.73 \pm 0.03$ mas for the object, putting it at a distance of 366$\pm$4 pc, i.e., inside the Orion star 
forming region.
Its magnitude, $G=14.45$, is too faint for \gaia\ to provide radial velocity measurements. Based on a limited set of APOGEE-2 spectra, \citet{Kounkel19} identified 
it as a double-lined binary, with a mass ratio of 0.988$\pm$0.063. It is thus a perfect target for binarity follow-up studies. As short-period binaries may be induced 
by the gravitational torque  from an additional companion \citep[e.g.,][]{2001ApJ...562.1012E,2014ApJ...793..137N}, it is also interesting to know if the system is 
in fact triple.

\gaia\  lists a slightly fainter ($\Delta G = 0.14$) and redder visual companion at a separation of  2.39$\arcsec$. With a parallax of $\varpi = 1.49 \pm 0.03$ mas, 
corresponding to a distance of about 670 pc, this optical companion is, however, unrelated to the much closer Ori OB1, as also confirmed by the analysis of its spectra 
presented in Sect.~\ref{Subsec:param}.
From now on, we will refer to this background star as CVSO~104\,B, while we will refer to our target as  CVSO~104\,A. 

The paper is organized as follows. Section~\ref{Sec:Observations} reports the observations used in this work. The results of the analysis of the data are then described 
in Section~\ref{Sec:Results} and discussed in Section~\ref{Sec:discussion}. We finally summarize our conclusions in Section~\ref{sect:Conclusions}.

\section{Observations}  
\label{Sec:Observations}

\subsection{Spectroscopy}
\label{Subsec:Spectroscopy}

High-resolution  spectroscopy ($R\simeq70,000$) was performed  with the Ultraviolet and Visual Echelle Spectrograph (UVES, \citealt{dekker00}) 
within the PENELLOPE Large Program (ESO Prog. ID. 106.20Z8; see Paper\,I).
Three UVES spectra were acquired within two days of the HST observation of CVSO~104\,A+B and a 
fourth one was purposely requested by us a few days later, after the clear detection of the two components in the spectra of this spectroscopic binary.
All these spectra were taken with the same position angle of 109\degree for the slit, aligning it with the optical pair.
Medium-resolution ($R\sim10,000-20,000$) broad-wavelength coverage spectroscopy was obtained using the X-Shooter instrument \citep{vernet11}. 
The strategy of spectroscopic observations and details on data reduction are explained in Paper\,I.
Particular attention was paid to the order trace and extraction to separate the spectra of the two stars of the visual pair. 

CVSO~104\,A+B was also observed during 5 visits (2 in February 2017, 3 in October 2017) in the framework of the APOGEE survey \citep{Ahumada20}, which 
obtains $R\sim22,500$ spectra in the $H$-band. 
Since the radial velocities measured on APOGEE spectra and reported in the SDSS Data Release 16 \citep{jonsson2020} were derived assuming the object 
as a single star, they are not usable. We therefore downloaded the APOGEE spectra\footnote{Available at \tt{http://skyserver.sdss.org/dr16/}} and derived 
the velocities of both components, in the same way as we did for the UVES and X-Shooter spectra (see Sec.~\ref{Subsec:RV}).

\begin{figure}
\begin{center}
\hspace{-.5cm}
\includegraphics[width=9.4cm,height=7.5cm]{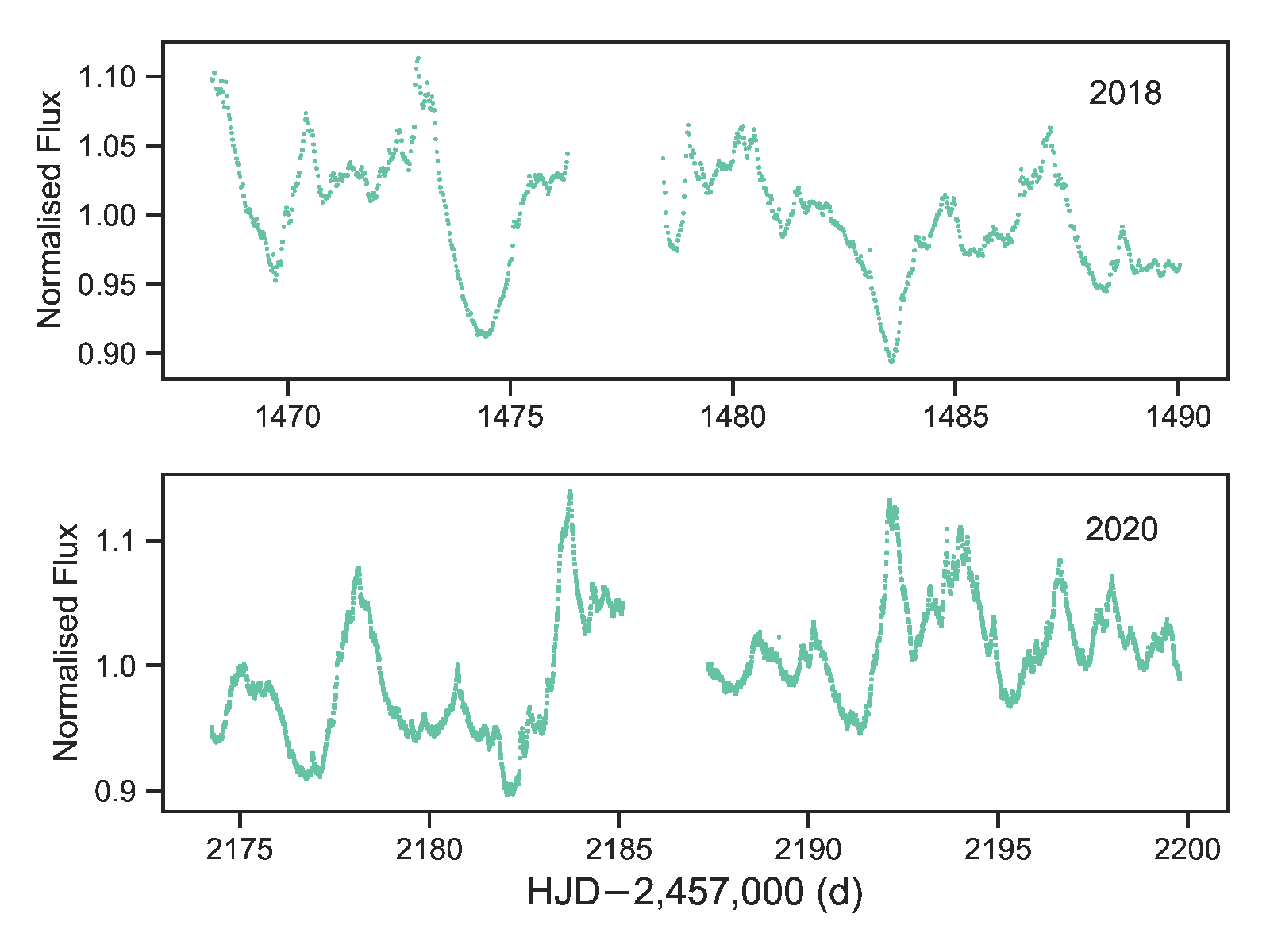}	
\vspace{-0.2cm}
\caption{TESS light curves of CVSO~104\,A+B in 2018 ({\it top}) and 2020 ({\it bottom}). }
\label{fig:TESS}
\end{center}
\end{figure}

\subsection{Photometry}
\label{Subsec:Obs_photo}

Space-born accurate photometry was obtained with  NASA's Transiting Exoplanet Survey Satellite (TESS; \citealt{Ricker15}). CVSO~104\,A+B was observed 
in sector 6 with 1426s exposure times between 2018-12-15 and 2019-01-06, as well as in sector 32 with 475s exposure times between 2020-11-20 and 
2020-12-16 -- this latter data set is contemporaneous to our spectroscopic observations. Given the pixel size of TESS, the optical companion will 
contribute to the light curve. We downloaded the data from the MAST archive and created light curves using the {\sc Lightkurve}\footnote{\tt{docs.lightkurve.org}} 
package by using a mask that contained all the neighboring pixels having S/N$>6$ (see Fig.~\ref{fig:TESS}). The 2018 data had a mean flux of 1244$\pm$51 e$^{-}$/s, 
while the 2020 data set had a similar mean flux of 1221$\pm$58 e$^{-}$/s.

To get the color information that is lacking in TESS data and to separate the contribution of the two visual components of the 
optical pair, several ground-based facilities were involved in the observation of this object during the HST and VST observations.
In the present work we will make use of data taken with four different instruments.

\begin{figure*}
\begin{center}
\hspace{-0.4cm}
\includegraphics[width=18.7cm]{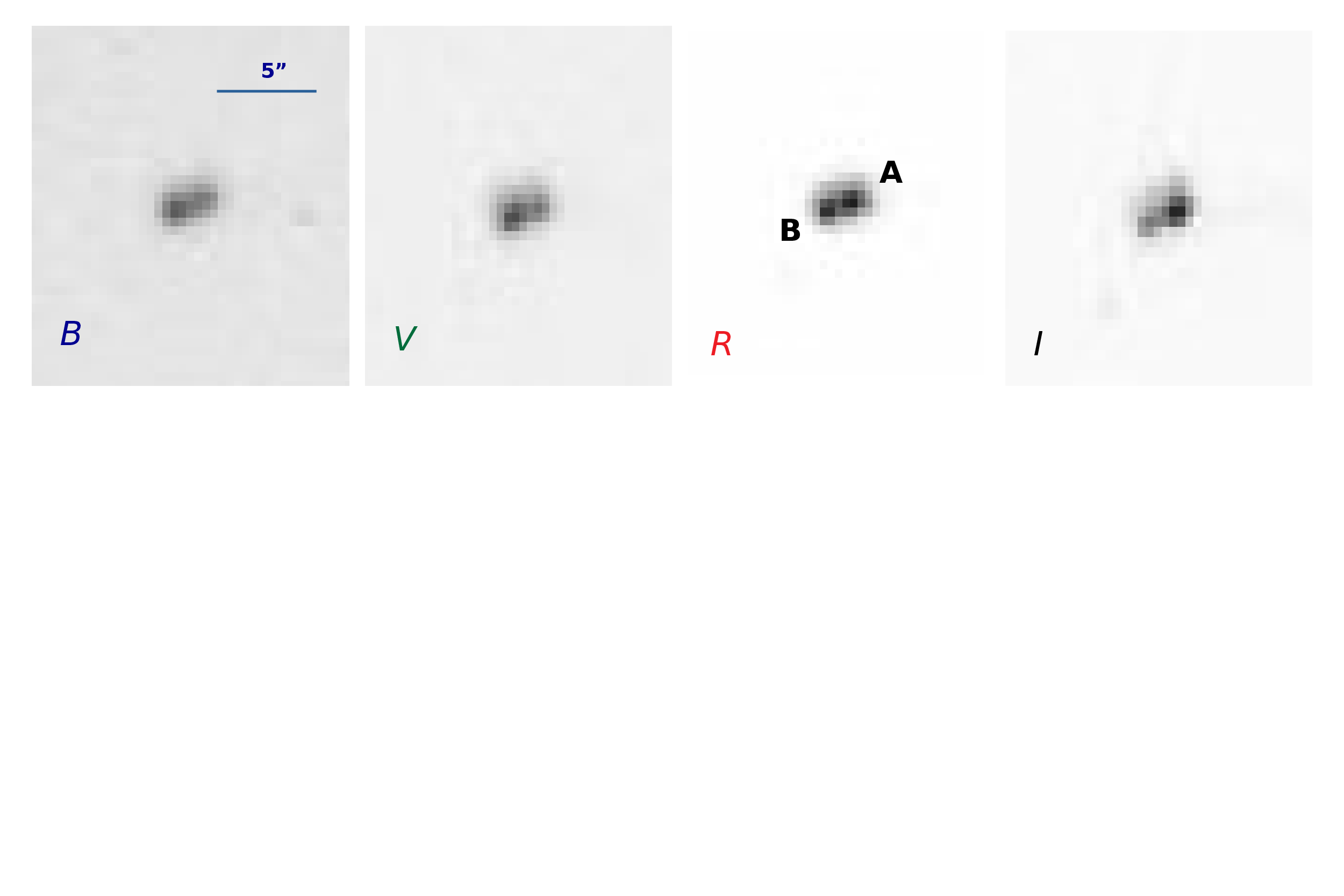}	
\vspace{.3cm}
\caption{Portion of images acquired at OACT on 16 Dec 2020 in $BVRI$ bands centered on the visual pair CVSO~104\,A+B. 
North is up and East to the left and the two components are indicated in the $R$-band image. 
Note how the relative brightness of the two stars changes in the different bands, with A becoming brighter than B in the $I$ band.
}
\label{Fig:images}
\end{center}
\end{figure*}

We observed CVSO~104 at the {\it M. G. Fracastoro} station (Serra La Nave, Mt. Etna,
1750 m a.s.l.) of the {\it Osservatorio Astrofisico di Catania} (OACT, Italy) from 25 November to 16 December 2020. 
We used the facility imaging camera at the 0.91\,m telescope with a set of broad-band Bessel filters ($B$, $V$, $R$, $I$, $Z$) as well  as two narrow-band 
H$\alpha$ filters centered on the line core ($H\alpha_9$) and on the redward continuum ($H\alpha_{18}$). 
The index $H\alpha_{18}$-$H\alpha_9$ is basically a measure of intensity of the H$\alpha$ emission in units of the continuum that can be converted into 
equivalent width (EW) of H$\alpha$ \citep{frasca2018}.
Details on photometric observations and data reduction can be found in Paper\,I.
We note that the two visual components are clearly resolved only in the nights with good seeing (see Fig.~\ref{Fig:images}), with the 
exception of the $Z$ filters in which the image is slightly out of focus.
We have therefore extracted the photometry of the stars in the field of CVSO~104 from the calibrated images using apertures of 5\arcsec\ of radius, which 
include both components of the visual pair. To measure the magnitude difference of the two visual components in the images taken with the best seeing, we have 
used an approach similar to that of \citet{Covino2004}.

CVSO\,104 was also observed from 6 November to 26 December, 2020 with the 0.8 m RC80 telescope of Konkoly Observatory (Hungary) using Bessel $BV$ and Sloan 
$r'i'$ filters. Details about the instrument, data reduction, and photometry are provided in \'Abrah\'am et al.~(in prep.). In most of the images the binary 
and the visual companion could be well separated. We performed PSF photometry separately for CVSO~104\,A and B in each image. 
As we will show in Sect.~\ref{Subsec:photo}, CVSO~104\,B is non-variable, therefore we calculated its average flux in each filter. Then we added these values 
to the fluxes of CVSO~104\,A, to obtain a full light curve of the visual pair CVSO~104\,A+B during the run to be used together with other data sets.  
To this aim the $r',i'$ magnitudes were converted to $R_C, I_C$ using the prescription given by 
Lupton (2005)\footnote{\scriptsize\tt http://www.sdss3.org/dr8/algorithms/sdssUBVRITransform.php}.

We used $BVri$ photometry collected by AAVSOnet\footnote{\scriptsize\tt https://www.aavso.org/aavsonet}, which is a set of robotic telescopes operated by 
volunteers for the American Association of Variable Star Observers (AAVSO). Stars in the AAVSO Photometric All-Sky Survey (APASS, \citealt{Henden2018}) 
were used to calibrate the AAVSO photometric data for our targets.
The $r,i$ magnitudes were converted to $R_C, I_C$ as was done for the Konkoly photometry.
We incorporated data taken by the amateur observers of the AAVSO\footnote{\scriptsize{Available at {\tt https://www.aavso.org/data-download}}}, in response to 
AAVSO Alert Notice 725.
Finally, some additional $BVRI$ photometry was obtained at the Crimean Astrophysical Observatory (CrAO) on the AZT-11  1.25~m telescope.

The photometry obtained with the latter instruments includes both components in the aperture. 
The multiband light curve of CVSO~104\,A+B during 30 nights including the ULLYSES and PENELLOPE campaigns is displayed in Fig.~\ref{fig:ground_LC}.

\section{Results}
\label{Sec:Results}

\subsection{Photometry}
\label{Subsec:photo}
The TESS light curves (Fig.~\ref{fig:TESS}) show rather stochastic variations, albeit quasi-periodic. 
The period analysis provides different results depending on the epoch of observations and the different techniques.
Periodograms reveal that the most likely period for the 2018 data set is 4.73 d, while for the 2020 data, this is 4.91 d, similar to the photometric variability 
period of 4.68 d in the $R$-band reported by \cite{Karim16}, as well as to the orbital period we derive (see below).
Using the CLEAN deconvolution algorithm \citep{Roberts1987}, we found the maximum power in 2018 at 2.31 d, i.e. about the half period of the periodogram and the 
second highest peak at about 4.53 d. In 2020, the highest peak corresponds to 4.73 d.  However, for both data sets, folding the data with any of the above periods 
does not reveal a convincing phase diagram. 

The ground-based multiband light curve (Fig.~\ref{fig:ground_LC}) shows a stochastic behaviour with at least two bursts fully observed in $BVRI$ bands. The intensity 
of the bursts is clearly stronger in the bluer bands. However, no intensification of the H$\alpha$ emission is observed in the OACT photometry contemporaneous to the 
stronger burst and the variations of H$\alpha$ EW do not seem to be correlated or anticorrelated with the brightness variations.

While the contribution of CVSO~104\,B is included in the photometric data presented in Fig.~\ref{fig:ground_LC}, our analyses suggest that this source is non-variable 
or that its variability is negligible in comparison to CVSO~104\,A. As a first check, we compared the \gaia\ EDR3 $G$-band magnitude uncertainty of component B with 
the corresponding \gaia\ magnitude uncertainties of other stars of the same brightness (we used Fig.\,5.15 
from the \gaia\ Early Data Release 3 Documentation V1.1\footnote{\tt gea.esac.esa.int/archive/documentation/GEDR3/} for this comparison). We concluded that the relative 
flux error of component B, 0.02\%, agrees with the representative numbers from similarly bright non-variable stars for which similar number of observations were taken. 
We also plotted the light curves of CVSO~104\,B using the spatially resolved photometry obtained at OACT and Konkoly observatories (Sect. \ref{Subsec:Obs_photo}). 
The plot shown in Fig.~\ref{fig:ground_compB} also confirms that the source was constant within the measurement uncertainties,
if we neglect very small systematic differences between Konkoly and OACT. In order to check this claim on more quantitative grounds, we calculated the $\chi^2$
and the probability $P(\chi^2)$ that the magnitude variations have a random occurrence \citep[e.g.,][]{Press1992}. We found, for the Konkoly dataset, values of 
$P(\chi^2)$ of 0.21, 0.22, 0.35, and 0.25 for $B$, $V$, $R_{\rm C}$, and $I_{\rm C}$ bands, respectively, which indicates these variations as non significant.

\subsection{Radial velocity}
\label{Subsec:RV}
The radial velocity was measured by cross-correlating the spectra with late-type templates. For the X-Shooter spectrum we selected only the regions with the best S/N 
in the VIS arm -- the one with the highest resolution -- and adopted as template a BT-Settl synthetic spectrum \citep{Allard2012} with [Fe/H]=0, \teff=4000\,K, 
and \logg=4.0. For the UVES spectra we adopted the same template and an HARPS spectrum of GJ\,514 (M1V, RV=14.606\,\kms, \citealt{jonsson2020}) obtaining the same 
results within the errors.
This analysis was carried out with the IRAF\footnote{IRAF is distributed by the National Optical Astronomy Observatory, which is operated by the Association of 
Universities for Research in Astronomy, Inc.} task {\sc fxcor} excluding emission lines and very broad features that can blur the peaks of the cross-correlation 
function (CCF). 
For a better measure of the centroids and full width at half maximum (FWHM) of the CCF peaks of the two components, we applied a two-Gaussian fit. 
The RV error, $\sigma_{\rm RV}$, was computed by {\sc fxcor} according to the fitted peak height and the antisymmetric noise as described by \citet{Tonry1979}.
We used the same BT-Settl template to measure RVs in the APOGEE spectra by cross-correlation. 
The individual values of RV measured in our and APOGEE spectra are listed in Table~\ref{Tab:RV}.

\setlength{\tabcolsep}{5pt}

\begin{table}	
\caption{Heliocentric radial velocities of the two components of CVSO~104\,A.}
\begin{tabular}{lrcrcl}
\hline
\hline
\noalign{\smallskip}
 HJD     &  RV$_{\rm a}$   &  $\sigma_{\rm RV_a}$ &  RV$_{\rm b}$   &  $\sigma_{\rm RV_b}$ & Instrument \\
(2\,400\,000+) &  \multicolumn{2}{c}{(\kms)} & \multicolumn{2}{c}{(\kms)} & \\
 \hline
\noalign{\smallskip}
 57794.66675 &  6.59    & 1.49 &  42.67 & 2.29   & APOGEE \\
 57795.66700 & $-10.39$ & 1.40 &  62.56 & 1.36   & APOGEE \\
 58033.96831 & 33.52    & 2.37 &  15.05 & 2.74   & APOGEE \\
 58037.98405 & $-8.64$  & 1.37 &  57.49 & 1.39   & APOGEE \\
 58039.00019 & 35.16    & 1.71 &   8.92 & 1.79   & APOGEE \\
59178.64280  & $-12.17$ & 0.68 &  63.68 & 0.56   & UVES \\
59179.62813  & 24.61    & 0.67 &  17.32 & 2.80   & UVES \\ 
59180.61009  & 61.23    & 0.19 & $-16.39$ & 0.23 & UVES \\
59180.63589  & 64.36    & 2.16 & $-14.54$ & 2.40 & XSHOO \\
59243.55343  & $-13.01$ & 0.61 &  64.25   & 0.59 & UVES \\
\noalign{\smallskip}
\hline \\
\end{tabular}
\label{Tab:RV}
\end{table}

\begin{figure}
\begin{center}
\hspace{-.5cm}
\includegraphics[width=9.cm]{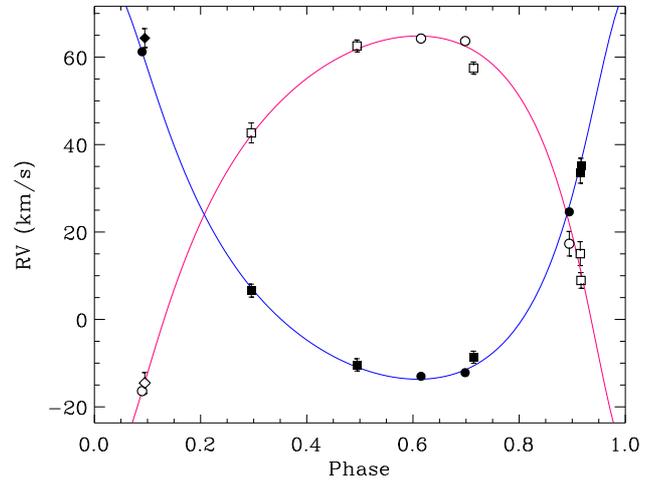}	
\vspace{0.cm}
\caption{Heliocentric radial velocity curve (circles = UVES, squares = APOGEE, diamonds = X-Shooter data) of CVSO~104\,A. Filled and open symbols have been used 
for the primary (more massive) and secondary component, respectively. The blue and red lines represent the orbital solution (Table\,\ref{Tab:Param}) 
for the primary and secondary component, respectively. 
}
\label{Fig:RV_curve}
\end{center}
\end{figure}

We searched for the best orbital period by applying periodogram analysis \citep{Scargle1982} and the CLEAN deconvolution algorithm \citep{Roberts1987} to
the RVs of the primary and secondary components.  
The data folded with the best period (maximum amplitude in the power spectrum) display a smooth variation with an asymmetrical shape, typical for an eccentric RV 
orbital motion. 
Then, we fitted the observed RV curve with the {\sc curvefit} routine \citep{Bevington} to determine the orbital parameters and their standard errors. 
This also allowed us to improve the determination of the orbital period, finding $P_{\rm orb}=5.025$ days  (see Fig.~\ref{Fig:RV_curve}). 
The orbital parameters are reported in Table~\ref{Tab:Param}.
We note that the high eccentricity, $e\simeq0.39$, is in line with the simulations of \citet{Zrake2021}, who show that equal-mass binaries accreting from a 
circumbinary disk evolve toward an orbital eccentricity of e$\sim$0.45, unless they start with a nearly circular orbit ($e\lesssim 0.08$). 

In addition, we searched for signatures of an eventual third companion, yet unresolved, in UVES spectra of CVSO~104\,A, independently on \gaia. For this purpose 
we applied the Broadening Function (BF) method \citep{Rucinski2012}, which is a linear deconvolution operation. Prior to the analysis, all emission lines were 
carefully removed from the spectra. However, no signatures of a third stellar body were found in the BF profiles. Furthermore, no residual peak at the velocity of 
the optical companion, CVSO~104\,B (RV\,$\simeq 42.9\pm0.5$\,\kms), was found either in the BFs or in the CCFs. This confirms that the extraction of the spectra 
allowed us to separate A and B components without significant contamination. 

\setlength{\tabcolsep}{2pt}

\begin{table}	
\caption{Orbital and stellar parameters of CVSO~104\,A.}
\begin{tabular}{lr|ccc}
\hline
\hline
\noalign{\smallskip}
\multicolumn{2}{c}{Orbital parameters}  &  \multicolumn{3}{c}{Stellar parameters} \\ 
\hline
\noalign{\smallskip}
 HJD0$^*$      &  788.156$\pm$0.023   &  & {\it Primary} & {\it Secondary}  \\
 $P_{\rm orb}$ ($d$) &  5.0253$\pm$0.0001 & SpT  &  M0   &  M2    \\
 $e$            &  0.395$\pm$0.048     & \teff (K) & 3770$\pm$180 & 3590$\pm$190 \\
 $\omega$ (\degree) &  160.9$\pm$1.9  & \vsini\ {\scriptsize(\kms)} &  7.5$\pm$1.0  & 6.0$\pm$1.0  \\
 $\gamma$ (\kms)    &  24.51$\pm$0.17  & $w_{\rm a,b}$    &  0.63$\pm$0.10 & 0.37$\pm$0.10 \\
 $K_{\rm a}$ (\kms)       &  45.93$\pm$0.27  & $W_{\rm Li}$ (m\AA) & 550$\pm$40 & 680$\pm$70 \\
 $K_{\rm b}$ (\kms)       &  48.85$\pm$0.27  & $A$(Li) & 3.2$\pm$0.2 & 3.6$\pm$0.3 \\
 $M_{\rm a}\sin^3i$ ($M_{\sun}$) & 0.185$\pm$0.013 & $R$ ($R_{\sun}$) & 1.07$\pm$0.02 & 0.99$\pm$0.02  \\
 $M_{\rm b}\sin^3i$ ($M_{\sun}$) & 0.171$\pm$0.012 & $L$ ($L_{\sun}$) & 0.21$\pm$0.04 & 0.15$\pm$0.04 \\
 $M_{\rm b}$/$M_{\rm a}$  &  0.921$\pm$0.008 & $M$ ($M_{\sun}$) & 0.57$\pm$0.15 & 0.43$\pm$0.15  \\
 $a\sin i$ ($R_{\sun}$) & 8.73$\pm$0.04  \\
\noalign{\smallskip}
\hline \\
\end{tabular}
\label{Tab:Param}
{\bf Notes.} $^{\rm *}$ Heliocentric Julian Date (HJD-2,457,000) of the periastron passage; the subscripts a and b refer to the primary (more massive) and 
secondary component, respectively.
\end{table}

\subsection{Stellar parameters}
\label{Subsec:param}

In the PENELLOPE survey we used the code \rotfit\  to determine the atmospheric parameters, \vsini, and veiling for single objects \citep[][Paper\,I]{Frasca2015,frasca2017}. 
We used this code for determining the parameters of CVSO~104\,B and found \teff$=5750\pm100$\,K, \logg$=4.40\pm0.12$\,dex, [Fe/H]$=0.08\pm0.07$\,dex, 
RV$=42.9\pm0.5$\,\kms, and \vsini$<2$\,\kms, i.e. CVSO~104\,B turns out to be a slowly-rotating Sun-like star. Furthermore, there is no lithium $\lambda 6708$\,\AA\ 
absorption line and no chromospheric emission is visible in the H$\alpha$ line core (see Fig.\,\ref{Fig:spe_B}). Therefore, we do not expect significant brightness 
variations, compared to CVSO~104\,A, from this background star, in agreement with the results from the ground-based photometry (Sect.~\ref{Subsec:photo}).

\begin{figure}
\begin{center}
\hspace{-.6cm}
\vspace{-0.2cm}
\includegraphics[width=8.8cm]{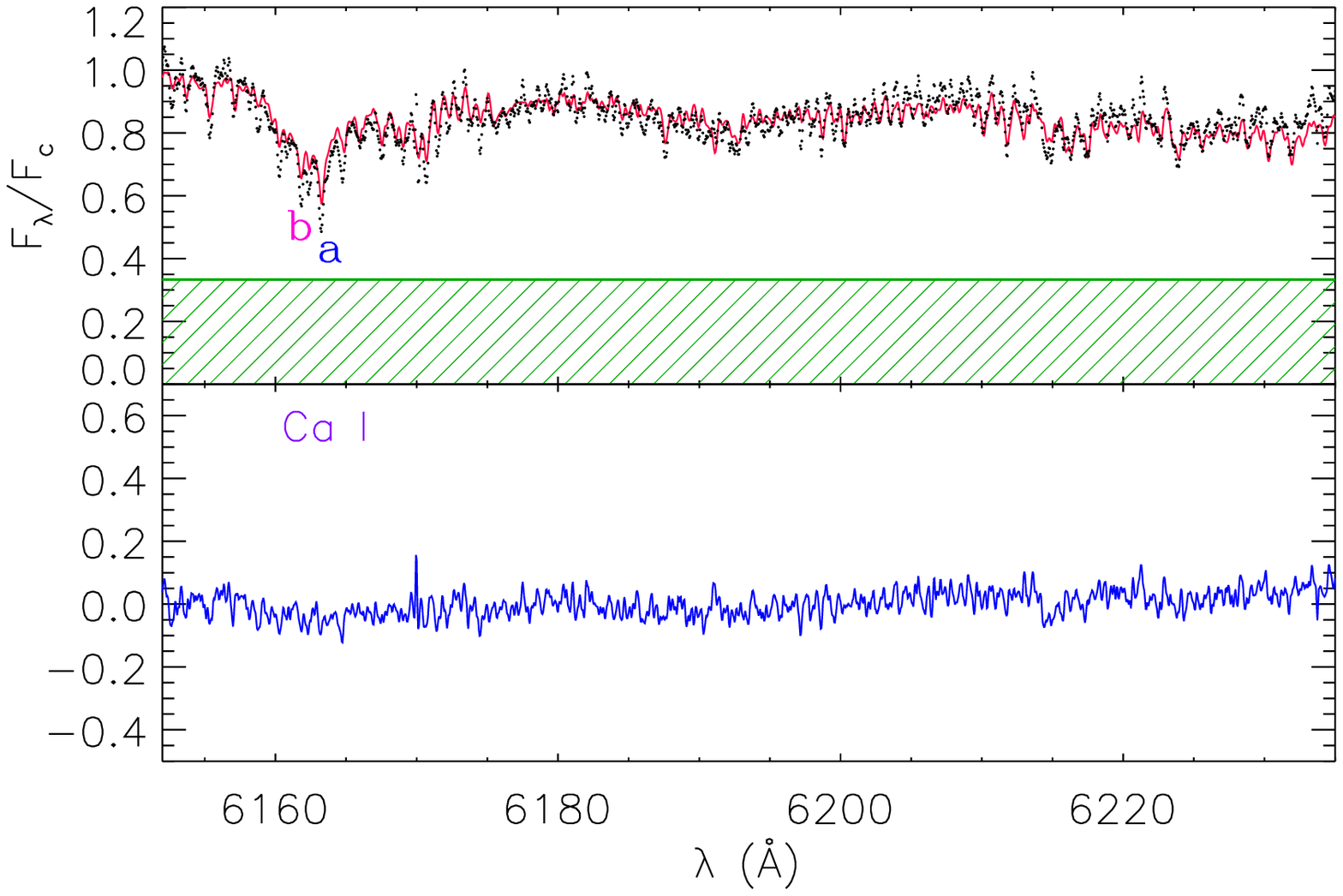}	
\vspace{-0.2cm}
\hspace{-.6cm}
\includegraphics[width=8.8cm]{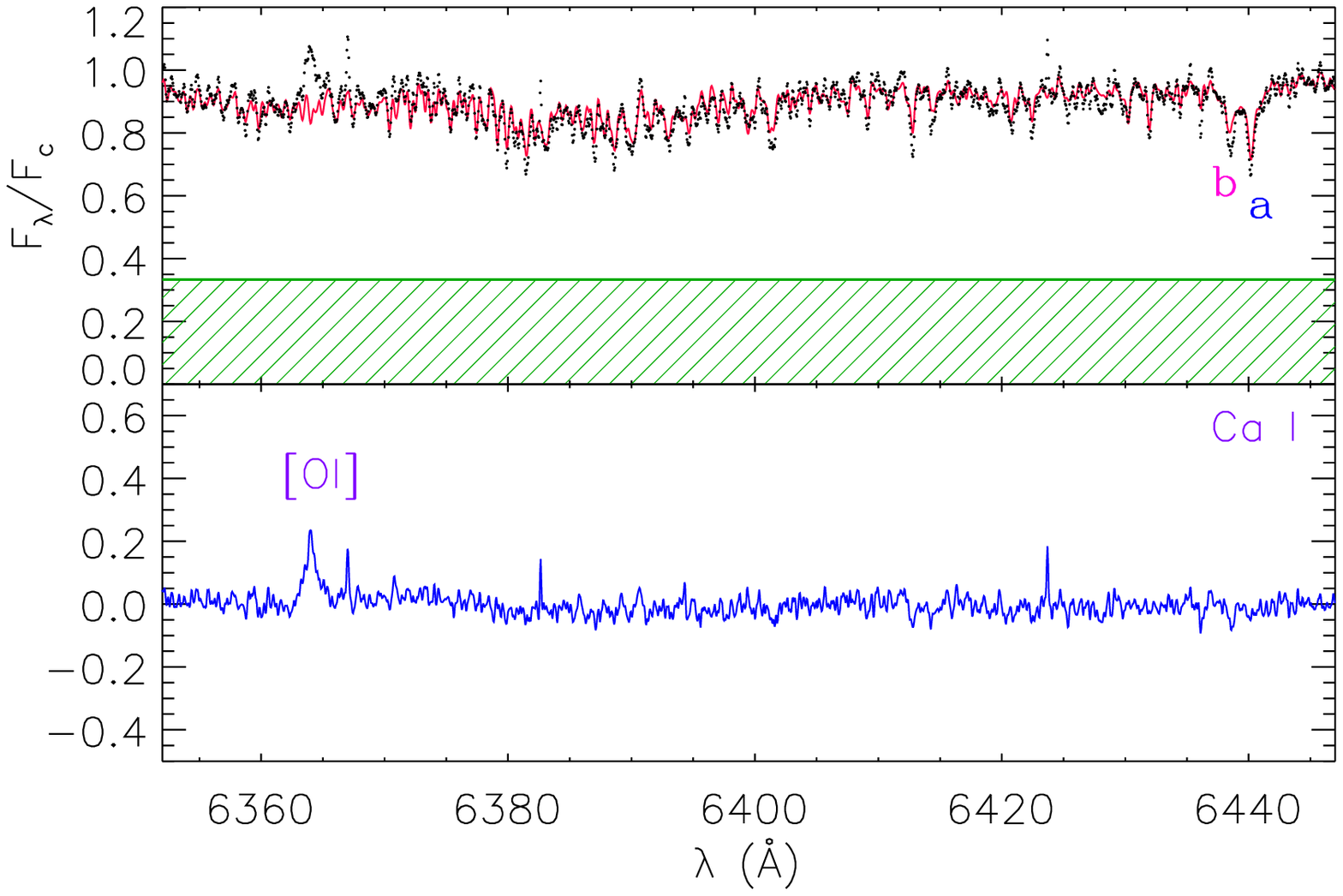}	
\vspace{0.cm}
\hspace{-.6cm}
\includegraphics[width=8.8cm]{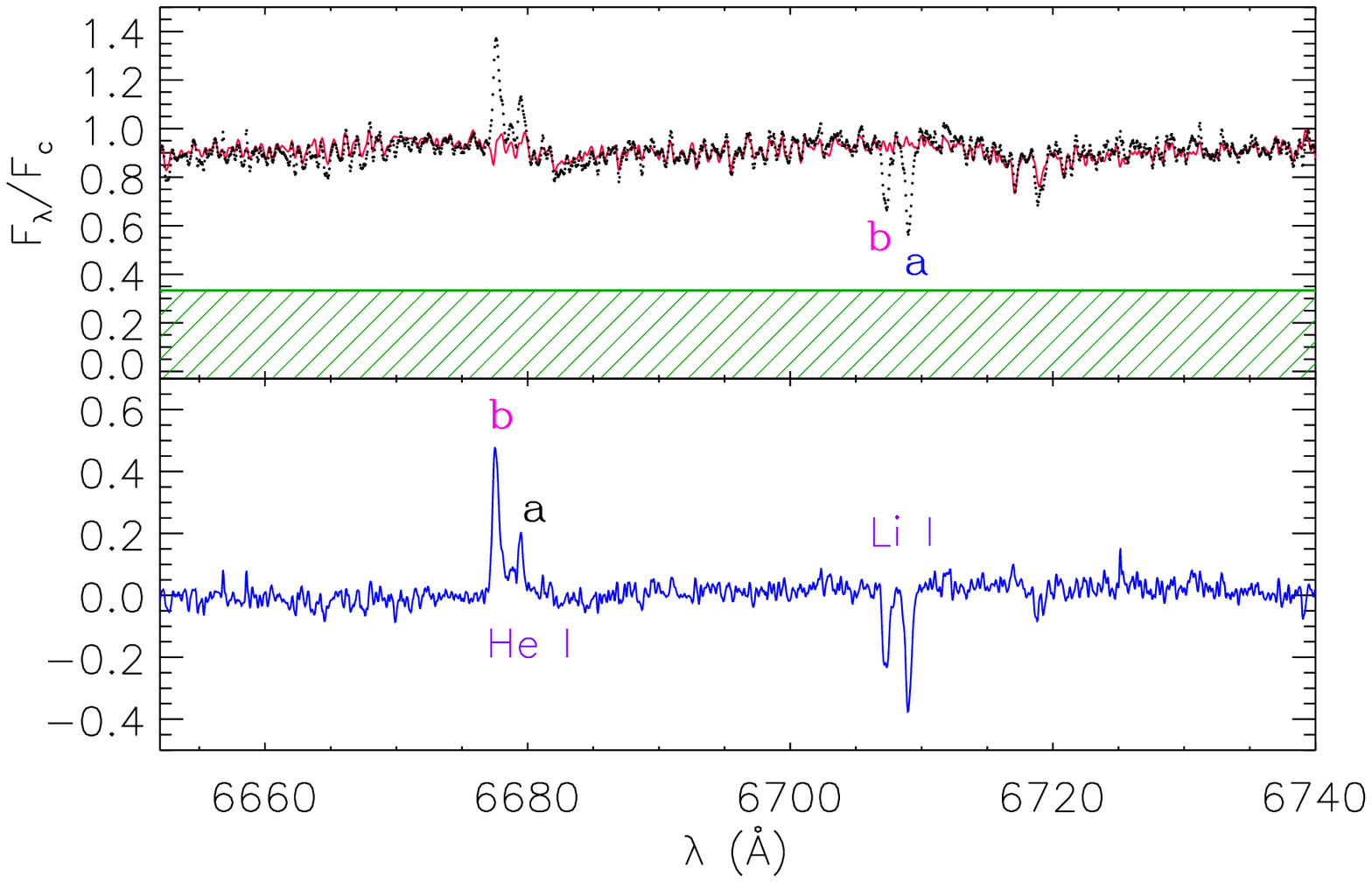}	
\caption{UVES spectrum of CVSO~104\,A (black dots) observed on JD\,=\,2\,459\,180 in three spectral regions around 6200 \AA\ ({\it top}), 6400 \AA\ ({\it middle}), and 
6700 \AA\ ({\it bottom}). In each box the synthetic spectrum, which is the weighted sum of two standard star spectra mimicking the primary and secondary component, 
is overplotted with a full red line. Prominent lines from the primary and secondary component are marked with a and b, respectively. The hatched green regions denote 
the level of the veiling. The residual spectra obtained by the subtraction of the photospheric template are displayed in the lower panel of each box with a blue line. 
}
\label{Fig:spe}
\end{center}
\end{figure}

In the case of double-lined spectroscopic binaries (SB2), we cannot use \rotfit, therefore we used \COMPO, a code developed in the 
{\sf IDL}\footnote{IDL (Interactive Data Language) is a registered trademark of  Harris Corporation.} 
environment by \citet{Frasca2006}, which has been adapted to the UVES spectra.
\COMPO\  adopts a grid of templates to reproduce the observed  spectrum, which is split into segments of 100 \AA\ each that are independently analyzed. 
As grid of templates, we used ELODIE spectra of 34 low-active slowly rotating stars with a spectral type in the range K2--M5.
The resolution of UVES was degraded to that of the ELODIE templates ($R=42,000$) by convolution with a Gaussian kernel with the proper width.
\COMPO\  does not derive the projected rotation velocities of the two components, which are instead estimated as \vsini$_{\rm a}=7.5$\,\kms\ and 
\vsini$_{\rm b}=6.0$\,\kms\ from the FWHM of the peaks of the CCF and kept as fixed parameters in the code. 
The RV separation of the two components is well-known from the CCF analysis and was used to build the composite ``synthetic'' spectrum. 
The flux ratio between the components, which has been expressed in terms of the flux contribution of the primary component in units of the continuum, $w_{\rm a}$, 
is instead let free to vary until a minimum $\chi^2$ is attained for each combination of spectra (1156 different combinations with the adopted grid). 
We note that the combination of two spectra with a relevant velocity separation reduces the intensity of the photospheric absorption lines of each 
component in a similar way as the veiling does.
However, no combination of late-type spectra was able to fairly fit the observed spectrum, unless we included a veiling. After several trials, we found as the best 
veiling $r=0.4$ for the UVES segments in the analyzed red region.
An example of the application of \COMPO\ is shown in Fig.~\ref{Fig:spe} for three spectral segments, around 6200\,\AA, 6400\,\AA\ and 6700\,\AA, of the UVES spectrum 
acquired on JD\,=\,2\,459\,180.

\begin{figure}
\begin{center}
\hspace{0cm}
\includegraphics[width=8.5cm]{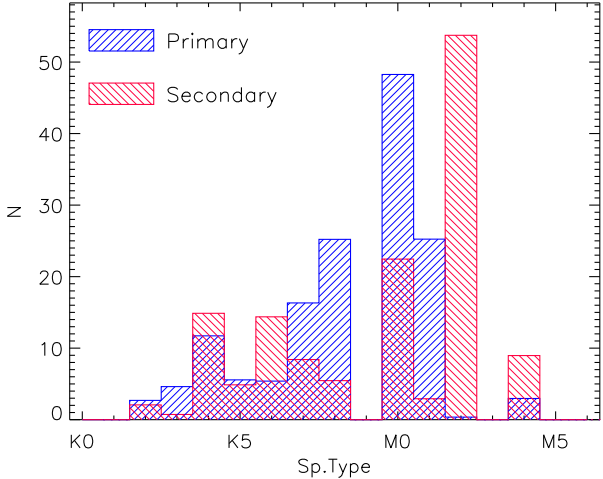}	
\vspace{-0.2cm}
\caption{Distribution of spectral types for the components of CVSO~104\,A.}
\label{fig:histo}
\end{center}
\end{figure}

To evaluate the atmospheric parameters (APs) and the flux contribution we kept only the best 25 combinations (based on the $\chi^2$), i.e. about the top 2\%, 
of primary and secondary spectra per each spectral segment and calculated the averages by weighting with the corresponding $\chi^2$. 
These parameters are listed in Table~\ref{Tab:Param}. 
The spectral types (SpT) of the components are taken as the mode of the spectral-type distributions (see Fig.~\ref{fig:histo}).

Equivalent widths of the \ion{Li}{i} 6708\AA\  line, $W_{\rm Li}$, for the  components of CVSO~104\,A have been measured on the residual spectrum (blue line in 
Fig.\ref{Fig:spe}). This procedure offers the advantage to remove any possible contamination from nearby iron lines.
The $W_{\rm Li}$ values were corrected for the veiling by multiplying them by $(1+r)$, adopting the value of veiling measured with \COMPO\ at red wavelengths $r=0.4$, 
and for the flux contribution to the composite spectrum by dividing them by $w_{\rm a}$ or $w_{\rm b}$ for the primary and secondary component, respectively.  
The error on the equivalent width was estimated as the product of the integration range and the mean error per spectral point, which results from the standard deviation 
of the flux values of the residual spectrum measured at the two sides of the line and is quoted in Table~\ref{Tab:Param} next to the $W_{\rm Li}$ values.
From these values we derived a Lithium abundance, $A$(Li), of 3.2$\pm$0.2 and 3.6$\pm$0.3 for the primary and secondary component, respectively, by using the curves 
of growth of \citet{Zapatero2002}. The abundance difference between primary and secondary component is within the uncertainties and is likely related to the uncertainties 
introduced by the correction for veiling and flux contribution.  These data are also reported in Table~\ref{Tab:Param}.

\subsection{Accretion and wind diagnostics}
\label{Subsec:accretion}

\begin{figure}
\begin{center}
\hspace{-0.4cm}
\includegraphics[width=9.2cm]{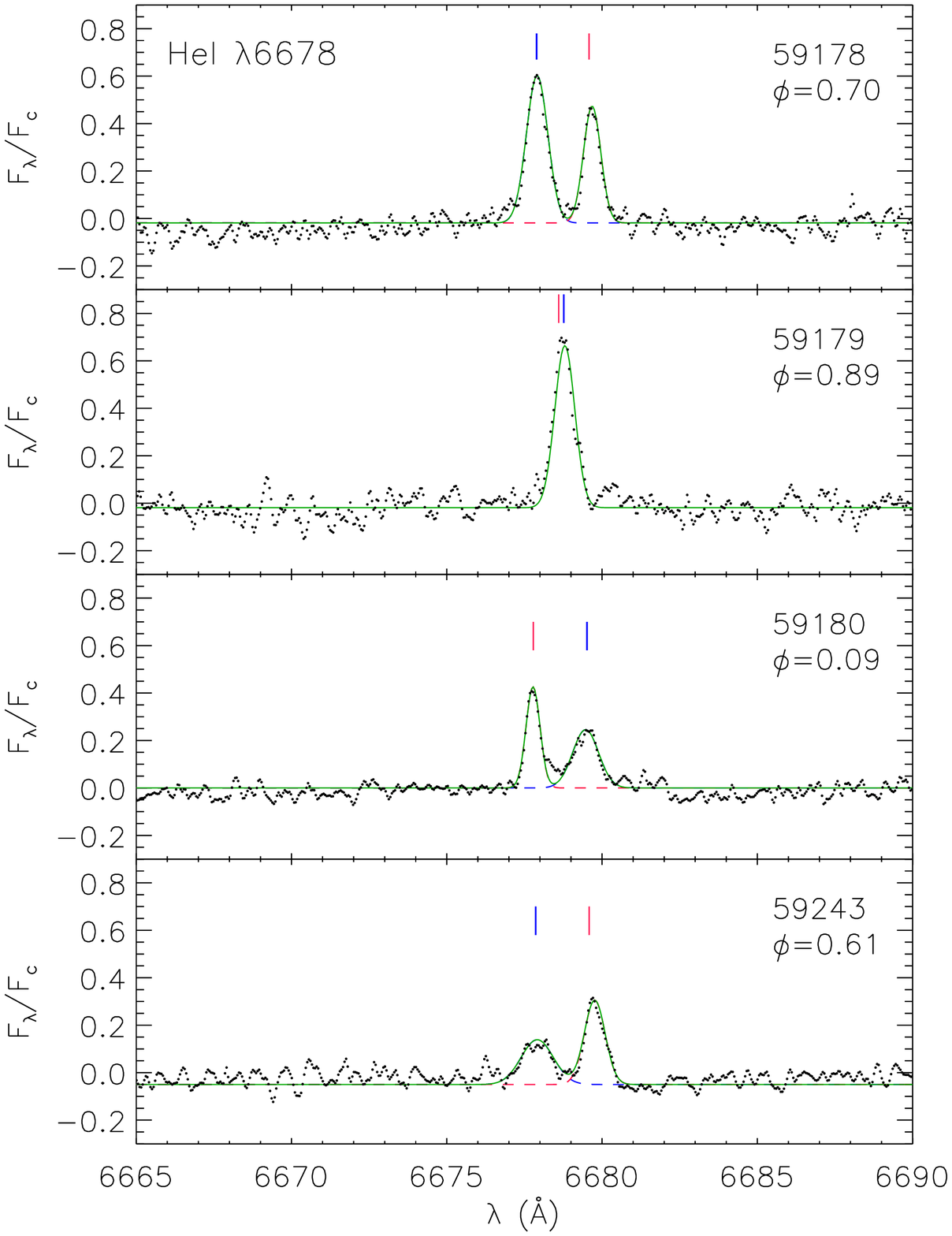} 
\vspace{-0.2cm}
\caption{Residual UVES spectra in the \ion{He}{i}\,$\lambda$6678\,\AA\ region (black dots). In each panel, the two-Gaussian fit to the emission peaks (green 
full line) and the individual Gaussians corresponding to the primary (blue dashed line) and secondary (red dashed line) component are overlaid. The blue and 
red vertical ticks mark the expected position of the lines of primary and secondary component, respectively, according to the photospheric RV. 
The reduced Julian day and the orbital phase ($\phi$) are marked in the upper right corner of each box. }
\label{fig:HeI6678}
\end{center}
\end{figure}

\begin{figure}
\begin{center}
\hspace{-0.3cm}
\includegraphics[width=9.2cm]{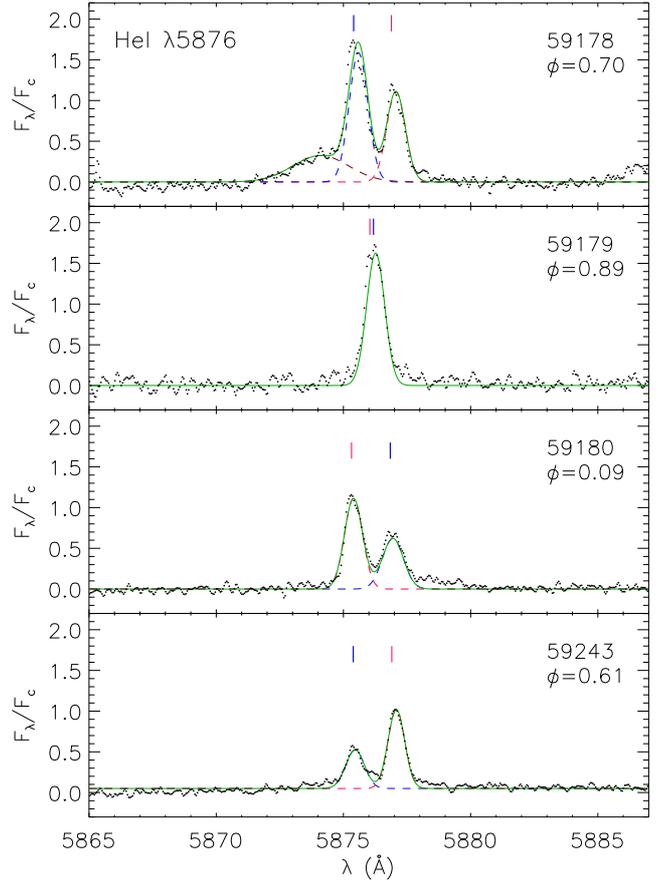}	
\vspace{-0.5cm}
\caption{Residual UVES  \ion{He}{i}\,$\lambda$5876\,\AA\ profiles (black dots). 
In each panel, the two-Gaussian fit to the emission peaks (green full line) and the individual Gaussians corresponding to the primary (blue dashed line) and 
secondary (red dashed line) component are overlaid. The blue and red vertical ticks mark the expected position of the lines of primary and secondary component, 
respectively, according to the photospheric RV.  
The dashed brown line in the upper panel represents the Gaussian fitted to the excess blueshifted emission observed only in this spectrum. The reduced Julian 
day and the orbital phase ($\phi$) are marked in the upper right corner of each box.}
\label{fig:HeI5876}
\end{center}

\end{figure}

The residual spectra are also very helpful to measure equivalent widths and fluxes of emission lines of the two components, when they are well separated in wavelength. 
We were successful to separate the contribution of the two components for \ion{He}{i} $\lambda$5876\,\AA\ and $\lambda$6678\,\AA\ lines in all the UVES spectra, with the 
exception of the spectrum taken at reduced Julian date RJD=59179, when the system was near a conjunction. We fitted two Gaussians to each spectrum to measure the 
centroid (RV), the FWHM, and equivalent width, $W_{line}$, of the line of each component (blue and red dashed lines in Figs.\,\ref{fig:HeI6678} and \ref{fig:HeI5876} 
for the primary and secondary component, respectively). 
We recover the line flux at Earth, $f_{\rm line}$, from the equivalent width $W_{\rm line}$ and the extinction corrected flux of the adjacent continuum,
namely $f_{\rm line} = W_{\rm line}f_{\rm cont}^0$ \citep[see, e.g.,][]{alcala17,frasca2018}, where the continuum flux is measured in the X-Shooter 
spectrum\footnote{A small correction was applied to the continuum flux based on the photometric variations. } and is corrected for the extinction by the factor 
$10^{0.4 A_{\lambda}}$, scaling $A_{\lambda}$ from the $A_V=0.35$\,mag (see Sect.~\ref{Subsec:SED}). The line luminosity is then derived adopting the distance $d$ as 
$L_{\rm line} =4\pi d^2 f_{\rm line}$.

The radial velocities, equivalent widths, and line luminosities 
are reported in Table\,\ref{Tab:emission_lines}, where we used the subscripts $a$ and $b$ for the quantities related to the primary and secondary component, respectively. 
A further broad emission feature, blueshifted with respect to the primary component, is visible in the \ion{He}{i}\,$\lambda$5876 line at RJD=59178. In this case we fitted 
the observed line profile with three Gaussians and report the parameters of this feature (brown dashed line in Fig.\,\ref{fig:HeI5876}) as $RV_3$, $W_3$, and 
$L_{\rm line}^3$ in Table\,\ref{Tab:emission_lines}. 

\begin{figure}
\begin{center}
\hspace{-0.6cm}
\includegraphics[width=9.2cm]{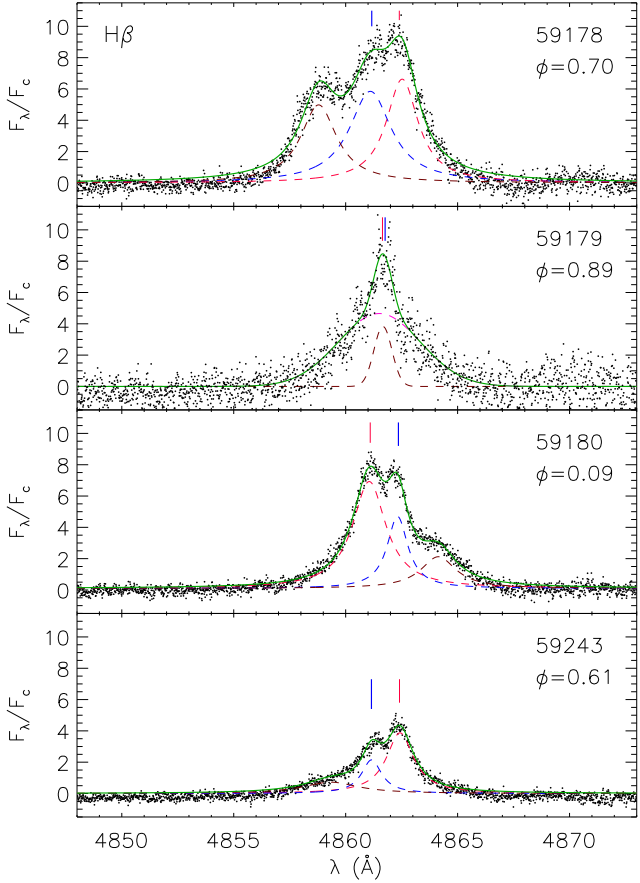}	
\vspace{0cm}
\caption{Residual UVES H$\beta$ profiles (black dots). The meaning of the symbols is as in Fig.\,\ref{fig:HeI5876}.}
\label{fig:Hbeta}
\end{center}
\end{figure}

The H$\beta$ emission profiles of the two components (Fig.~\ref{fig:Hbeta}) are much wider than the \ion{He}{i} lines and, as a consequence, they are strongly blended. 
However, we could still deblend them with multi-component fits of Lorentzian profiles. A third emission component is always visible out of the conjunction as a blueshifted 
or redshifted feature, depending on the orbital phase. It displays the maximum intensity in the first epoch (RJD=59178), when we also observed the excess \ion{He}{i} emission.
All the three H$\beta$ emission peaks display a much smaller intensity at RJD=59243, although the system configuration is the same as the first epoch of UVES observations, 
suggesting an intrinsic accretion variation in addition to any potential geometric effect. The RV, equivalent width and line luminosity of the H$\beta$ emission components 
are also quoted in Table\,\ref{Tab:emission_lines}.

We cannot distinguish the emission contribution of the two stars in the H$\alpha$ line (Fig.~\ref{fig:Halpha}), but we always see the blueshifted and redshifted excess 
component at a velocity similar to that of H$\beta$. A two-Gaussian fit has allowed us to separate the contribution of the latter feature from the integrated emission 
coming from the two stars, which we also report in Table\,\ref{Tab:emission_lines}.  
We think that this broad excess emission feature can be related to the complex structure of accretion funnels. 

The strong intensity variation from the first to the last spectrum is also observed in H$\alpha$ and suggests a variable accretion rate.

We have calculated the accretion luminosity, $L_{\rm acc}$, with the $\log L_{\rm acc}$--$\log L_{\rm line}$ linear relations proposed by \citet{alcala17}.
The mass accretion rate, \Macc, was then derived from \Lacc\ according to 
\begin{equation}
  \dot{M}_{\rm acc} = ( 1 - \frac{R_{\star}}{R_{\rm in}} )^{-1} ~ \frac{L_{\rm acc} R_{\star}}{G M_{\star}}, 
\end{equation}
\noindent{where  $R_{\star}$ and $R_{\rm in}$ are the stellar radius and inner-disk radius (assumed to be $R_{\rm in}=5R_{\star}$), respectively \citep[see][]{Gullbring1998,Hartmann1998}.}
At each epoch, we have calculated the mean values of \Lacc\ and \Macc\ (reported in Table~\ref{Tab:accretion}), which have been  obtained averaging the individual values 
derived from the two \ion{He}{i} and H$\beta$ lines. The errors include the relative errors of the line luminosities and the standard deviation of \Lacc\ values from the 
three diagnostics.

\setlength{\tabcolsep}{3pt}

\begin{table*}	
\caption{Multi-component analysis of permitted emission lines of CVSO~104\,A in the UVES spectra.}
\begin{tabular}{lrrrrrrrrrrccc}
\hline
\hline
\noalign{\smallskip} 
 RJD  & RV$_a$ & RV$_b$ & RV$_3$ & FWHM$_a$  & FWHM$_b$  & FWHM$_3$  & $W_a$ & $W_b$ & $W_3$ & $L_{line}^a$ & $L_{line}^b$  & $L_{line}^3$ \\ 
     & \multicolumn{3}{c}{~~~~~~~~~(\kms)}  & \multicolumn{3}{c}{~~~~~~~~~(\kms)} & \multicolumn{3}{c}{~~~(\AA)} & \multicolumn{3}{c}{($10^{-6}L_{\sun}$)} \\ 
 \hline
\noalign{\smallskip}
 \multicolumn{12}{c}{\bf \ion{He}{i}\,$\lambda$6678} \\
59178  &  -10.9$\pm$1.9 & 68.6$\pm$2.2 & \dots & 38$\pm$5 & 30$\pm$5 & \dots & 0.48$\pm$0.07 & 0.30$\pm$0.06 & \dots & 7.9$\pm$1.1 & 5.1$\pm$1.0 & \dots \\ 
59179  &  \multicolumn{2}{c}{26.9$\pm$1.9} & \dots &  \multicolumn{2}{c}{36$\pm$5} & \dots & \multicolumn{2}{c}{0.59$\pm$0.09} & \dots &  \multicolumn{2}{c}{8.8$\pm$1.5} & \dots \\ 
59180  &  60.1$\pm$5.1 & -16.8$\pm$2.1 & \dots & 57$\pm$13 & 26$\pm$5 & \dots & 0.25$\pm$0.05 & 0.24$\pm$0.03 & \dots & 4.2$\pm$0.8 & 3.9$\pm$0.5 & \dots \\ 
59243  & -12.6$\pm$7.7  &  73.1$\pm$3.0 & \dots & 43$\pm$18 & 30$\pm$7 & & 0.22$\pm$0.05 & 0.31$\pm$0.04 & \dots & 3.6$\pm$0.8 & 6.1$\pm$0.8 & \dots \\ 

\noalign{\smallskip}
\hline 
\noalign{\smallskip}
 \multicolumn{12}{c}{\bf \ion{He}{i}\,$\lambda$5876} \\
59178  & -7.6$\pm$1.2 &  70.2$\pm$1.6 & -82.4$\pm$21.4 & 46$\pm$4 &  45$\pm$4 & 124$\pm$49 &  1.57$\pm$0.08 & 1.07$\pm$0.08 & 0.76$\pm$0.15 & 16.3$\pm$0.8 & 11.2$\pm$0.8 & 7.8$\pm$1.5 \\ 
59179  & \multicolumn{2}{c}{29.5$\pm$1.1} & \dots &  & & &  \multicolumn{2}{c}{1.76$\pm$0.14} & \dots & \multicolumn{2}{c}{16.6$\pm$1.4} &  \dots \\ 
59180  & 65.3$\pm$2.6 & -12.5$\pm$1.3 & \dots & 58$\pm$7 &  40$\pm$3 & \dots  & 0.86$\pm$0.09 & 0.90$\pm$0.08 & \dots & 9.0$\pm$0.9 & 9.3$\pm$0.8 & \dots \\ 
59243  & -9.7$\pm$3.2 &  73.7$\pm$1.5 & \dots &  49$\pm$8 & 41$\pm$4 & \dots &  0.40$\pm$0.07 & 0.66$\pm$0.06 & \dots & 4.1$\pm$0.7 & 6.9$\pm$0.6 & \dots \\ 
\noalign{\smallskip}
\hline 
\noalign{\smallskip}
 \multicolumn{12}{c}{\bf H$\beta$} \\
59178  & -16.3$\pm$3.3 &  71.5$\pm$2.8 &  -158.5$\pm$3.6 & 120$\pm$8 & 111$\pm$4 & 123$\pm$5  &  22.3$\pm$1.8 & 24.9$\pm$1.7 & 18.9$\pm$1.9 & 214.5$\pm$17.3 & 239.5$\pm$16.3 & 182.0$\pm$18.3 \\ 
59179  &  \multicolumn{2}{c}{18.7$\pm$2.1} & \dots  &  \multicolumn{2}{c}{190$\pm$40} & \dots  &  \multicolumn{2}{c}{21.2$\pm$2.0}  & \dots  & \multicolumn{2}{c}{163.1$\pm$19.2}  & \dots \\ 
59180  &  59.4$\pm$2.6 & -19.3$\pm$3.3 &  171.9$\pm$4.7 & 69$\pm$5 & 116$\pm$3 & 116$\pm$8 & 8.0$\pm$0.9 & 19.1$\pm$1.6 & 6.1$\pm$1.2 & 77.4$\pm$8.7 & 183.8$\pm$15.4 & 58.6$\pm$11.5 \\ 
59243  & -12.7$\pm$3.5 &  65.6$\pm$2.3 & -124.8$\pm$18.7 & 70$\pm$13 &  78$\pm$6 & 285$\pm$32 &  6.7$\pm$0.7 & 10.8$\pm$0.9 & 3.6$\pm$1.0 & 64.9$\pm$6.8 & 103.9$\pm$8.7 & 34.8$\pm$9.7 \\ 
\noalign{\smallskip}
\hline 
\noalign{\smallskip}
 \multicolumn{12}{c}{\bf H$\alpha$} \\
59178  &  \multicolumn{2}{c}{47.1$\pm$3.7} & -168.8$\pm$5.3 & \multicolumn{2}{c}{200$\pm$12} & 150$\pm$15 &  \multicolumn{2}{c}{60.0$\pm$1.5} & 26.8$\pm$1.1 & \multicolumn{2}{c}{1177$\pm$29} & 525$\pm$22 \\ 
59179  &  \multicolumn{2}{c}{24.1$\pm$3.8} &  \dots & \multicolumn{2}{c}{250$\pm$19} & \dots &  \multicolumn{2}{c}{50.7$\pm$1.7} & \dots  & \multicolumn{2}{c}{895$\pm$33}  & \dots \\ 
59180  &  \multicolumn{2}{c}{21.6$\pm$3.3} &  210.8$\pm$9.0 & \multicolumn{2}{c}{188$\pm$10} & 131$\pm$21 &  \multicolumn{2}{c}{59.8$\pm$1.0} & 9.9$\pm$0.9 & \multicolumn{2}{c}{1174$\pm$20} & 194$\pm$18 \\ 
59243  &  \multicolumn{2}{c}{51.4$\pm$6.6} &  -86.2$\pm$14.5 & \multicolumn{2}{c}{152$\pm$11} &  214$\pm$18 &  \multicolumn{2}{c}{21.3$\pm$1.0} & 14.5$\pm$1.2 & \multicolumn{2}{c}{419$\pm$20} & 284$\pm$24 \\ 
\hline 
\end{tabular}
\label{Tab:emission_lines}
\end{table*}


\begin{table}	
\caption{Accretion luminosity and mass accretion rates of the components of CVSO~104\,A.}
\begin{tabular}{lcccc}
\hline
\hline
\noalign{\smallskip} 
 RJD  & $\log L_{\rm acc}^a$ & $\log L_{\rm acc}^b$ & $\log\dot{M}_{\rm acc}^a$ & $\log\dot{M}_{\rm acc}^b$\\
     &  \multicolumn{2}{c}{($L_{\sun}$)} & \multicolumn{2}{c}{($M_{\sun} yr^{-1}$)} \\
 \hline
\noalign{\smallskip}
59178   & $-1.70\pm$0.18 & $-1.8\pm$0.30 & $-8.82\pm$0.18 & $-8.86\pm$0.30 \\
59179   & \multicolumn{2}{c}{$-1.45\pm$0.22} & \multicolumn{2}{c}{\dots} \\ 
59180   &  $-2.08\pm$0.24 & $-1.95\pm$0.31 & $-9.20\pm$0.24 & $-8.98\pm$0.31 \\
59243   &  $-2.27\pm$0.28 & $-2.01\pm$0.29 & $-9.39\pm$0.28 & $-9.04\pm$0.29 \\
\noalign{\smallskip}
\hline 
\noalign{\smallskip}
\end{tabular}
\label{Tab:accretion}
\end{table}

\begin{figure}
\begin{center}
\hspace{-0.6cm}
\includegraphics[width=9.2cm]{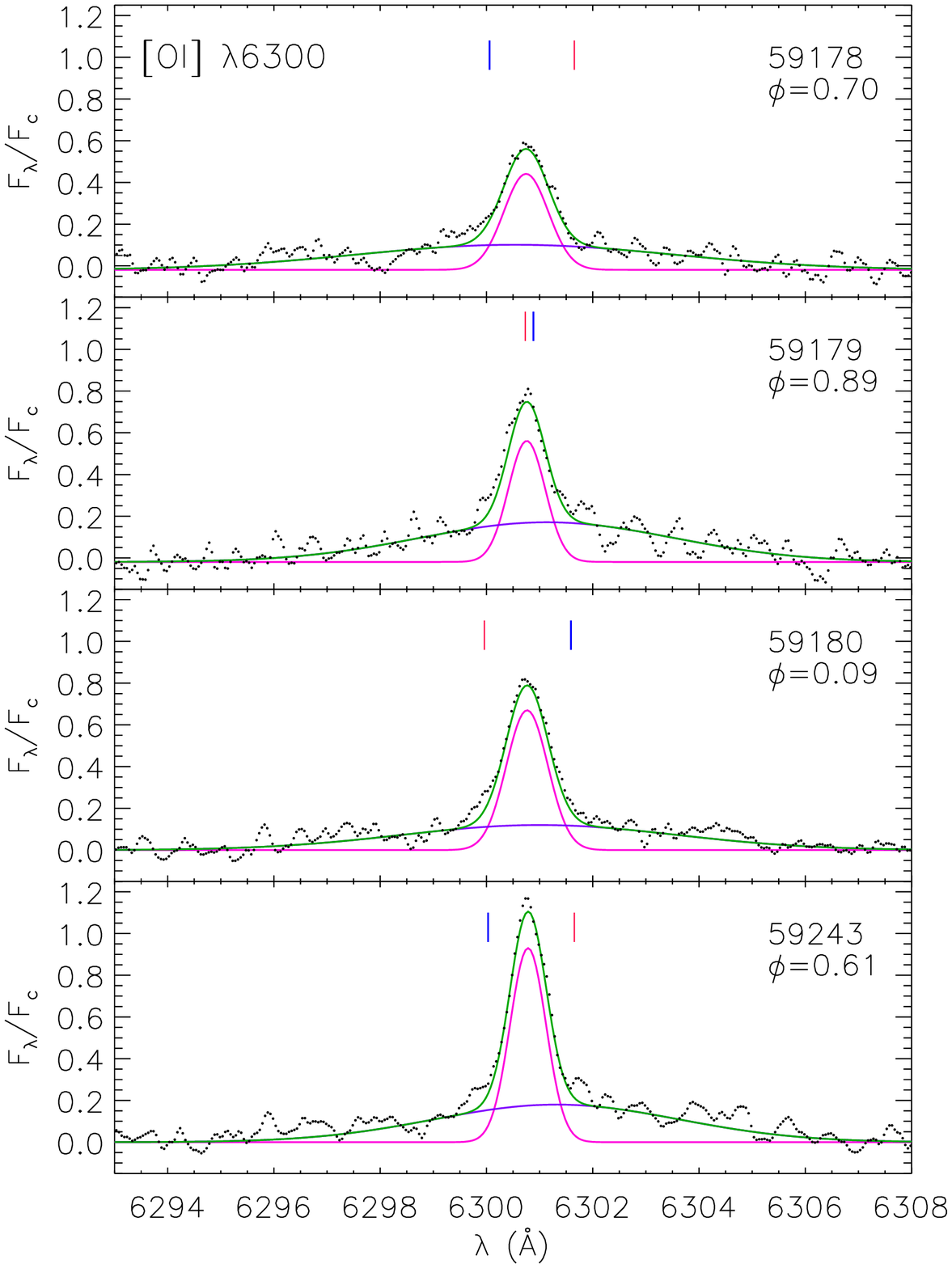}	
\vspace{0cm}
\caption{Residual UVES spectra in the [\ion{O}{i}]\,$\lambda6300$ region (black dots). The blue and red vertical ticks mark the expected position of the lines of 
primary and secondary component, respectively, according to the photospheric RV. The Gaussian fits to the broad and narrow component of the line profile are 
overplotted with purple and magenta lines, respectively, while their sum is plotted with a green line. The reduced Julian day and the orbital phase ($\phi$) are 
marked in the upper right corner of each box. }
\label{fig:OI}
\end{center}
\end{figure}

The [\ion{O}{i}]\,$\lambda$6300\,\AA\ line is always observed as a single and rather symmetric feature (see Fig.\,\ref{fig:OI}). Its radial velocity spans from 
+19 to +26 \kms, therefore it is always close to the barycentric velocity. Its equivalent width in the UVES spectra does not change very much, being about 0.80, 
0.86, 1.09, and 1.20\,\AA, from the first to the last epoch of UVES observations. Interestingly, it is slightly stronger when the permitted lines are weaker. 
The [\ion{O}{i}]\,$\lambda$6363\AA\ line displays a similar behavior. This does not necessarily mean a real intensity variation, but it could be instead the result 
of a decrease of the excess continuum flux due to the accretion, viz., the veiling.

The observed profile of the [\ion{O}{i}]\,$\lambda$6300\,\AA\ line displays broad wings that cannot be reproduced with a single Gaussian. We have therefore 
interpreted them as the superposition of a narrow and a broad component, which have been fitted with Gaussians. The broad component is normally associated with 
magneto-hydrodynamic winds from the inner ($<1$ au) disk, while the narrow one is related to photoevaporated winds originating in a much more extended region 
\citep[up to 100 au, e.g.][]{Ercolano17}.
The parameters of the narrow (N) and broad (B) emission component are reported in Table~\ref{Tab:OI}. The  narrow emission component has an average width of 
FWHM$\simeq$\,44\,\kms, while it is about 288 \kms\ for the broad one.
Following the prescriptions applied in different studies of forbidden emission lines \citep{Simon16,McGinnis18,Fang18,Banzatti19,Gangi20} we can estimate the 
emission size of these components under the assumption that the line widths are dominated by Keplerian broadening.
Assuming an inclination of $i=43$\,\degree\ and a total mass of 1\,$M_{\sun}$ (Sect.~\ref{Subsec:SED}), the narrow component should be emitted by a region of 
$\approx$ 0.84 au (180 $R_{\sun}$), which is compatible with a circumbinary disk, as also found from the SED analysis.
The size of the source of the broad component should be $\approx$\,0.02 au (4.3\,$R_{\sun}$).

\begin{table}	
\caption{Multicomponent analysis of the [\ion{O}{i}]\,$\lambda$6300\,\AA\ emission line.}
\begin{tabular}{lcccccc}
\hline
\hline
\noalign{\smallskip} 
 RJD  & RV$_N$ & FWHM$_N$  & $I_N$ & RV$_B$ & FWHM$_B$  & $I_B$  \\
     &  \multicolumn{2}{c}{(\kms)} &  & \multicolumn{2}{c}{(\kms)}  \\
 \hline
\noalign{\smallskip}
59178  & 21.0$\pm$3.7 & 46.9$\pm$7.8 & 0.46(6) & 11.1$\pm$18.9 & 355$\pm$84 & 0.12(2) \\
59179  & 21.6$\pm$3.3 & 38.7$\pm$6.9 & 0.58(8) & 39.6$\pm$12.6 & 267$\pm$51 & 0.19(3) \\ 
59180  & 22.0$\pm$2.0 & 43.8$\pm$4.0 & 0.67(5) & 32.8$\pm$13.4 & 289$\pm$50 & 0.12(2) \\
59243  & 22.8$\pm$1.6 & 37.9$\pm$3.4 & 0.93(8) & 48.3$\pm$10.8 & 258$\pm$43 & 0.18(3) \\
\noalign{\smallskip}
\hline 
\noalign{\smallskip}
\end{tabular}
{\bf Notes.} $I_N$ and $I_B$ are the intensities of the narrow and broad component, respectively, in units of the continuum.
\label{Tab:OI}
\end{table}

\subsection{Spectral energy distribution}
\label{Subsec:SED}

\begin{figure}
\begin{center}
\hspace{0cm}
\includegraphics[width=9cm]{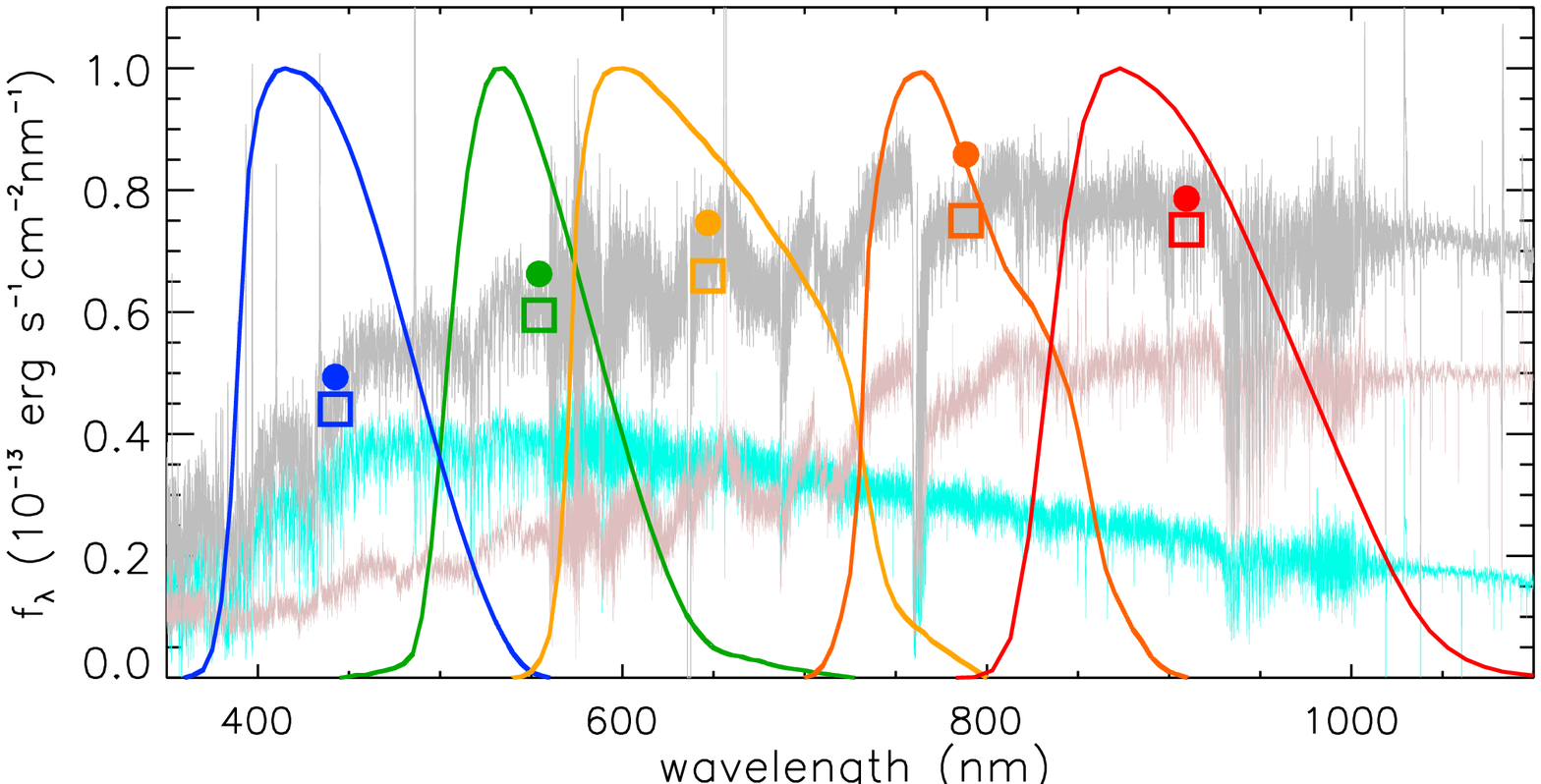}	
\includegraphics[width=9cm]{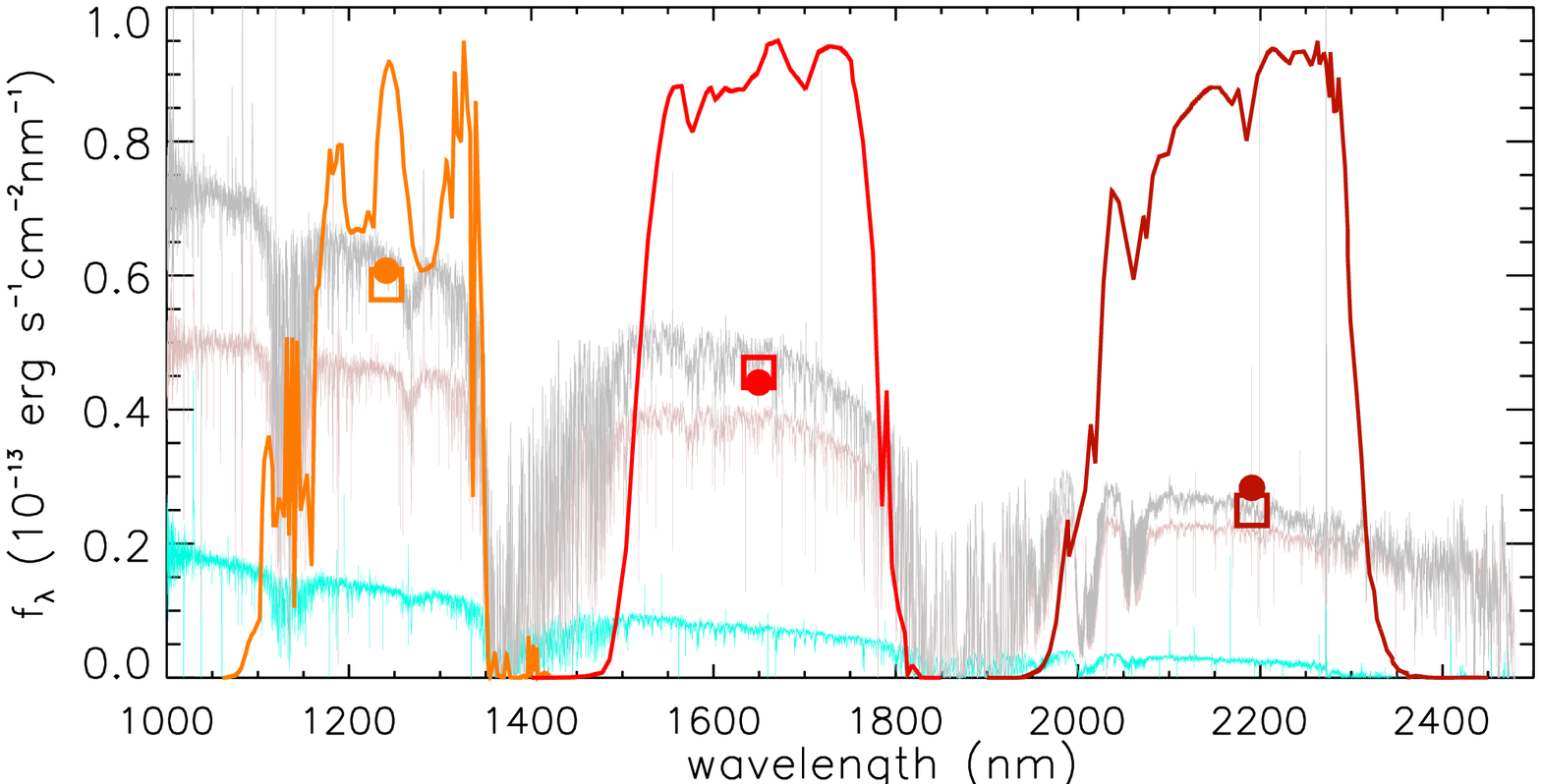}	
\vspace{0cm}
\caption{X-Shooter flux-calibrated spectrum of CVSO~104\,A (pink line), CVSO~104\,B (cyan line), and the sum of the two (grey line) in the visual ({\it top panel}) 
and NIR region ({\it bottom panel}). The contemporaneous photometry obtained from OACT in the $BVR_{\rm C}I_{\rm C}z'$ bands and the 2MASS $JHK_{\rm s}$ photometry 
are displayed with colored dots. The filter bandpasses and the synthetic photometry (open squares) obtained integrating the combined spectrum (A+B) over these 
bandpasses are overlaid with the same color as the photometric points.}
\label{fig:Xsho}
\end{center}
\end{figure}

To study the shape of the spectral energy distribution (SED) of CVSO~104\,A, excluding the contribution of the optical companion, we used both the OACT photometry 
and synthetic photometry made on the flux-calibrated X-Shooter spectra of the two stars (see Fig.~\ref{fig:Xsho}).
 We extended the SED to mid-infrared (MIR) and far-infrared (FIR) 
wavelengths by adding flux values from the literature. These data are quoted in Table~\ref{Tab:SED}.  As the contribution of the visual companion B is strongly 
decreasing at longer wavelengths (see Fig.~\ref{fig:Xsho}), we assign the MIR (WISE) and FIR ({\it Herschel}) fluxes to the component A.

\begin{figure}[ht]
\includegraphics[width=9.2cm]{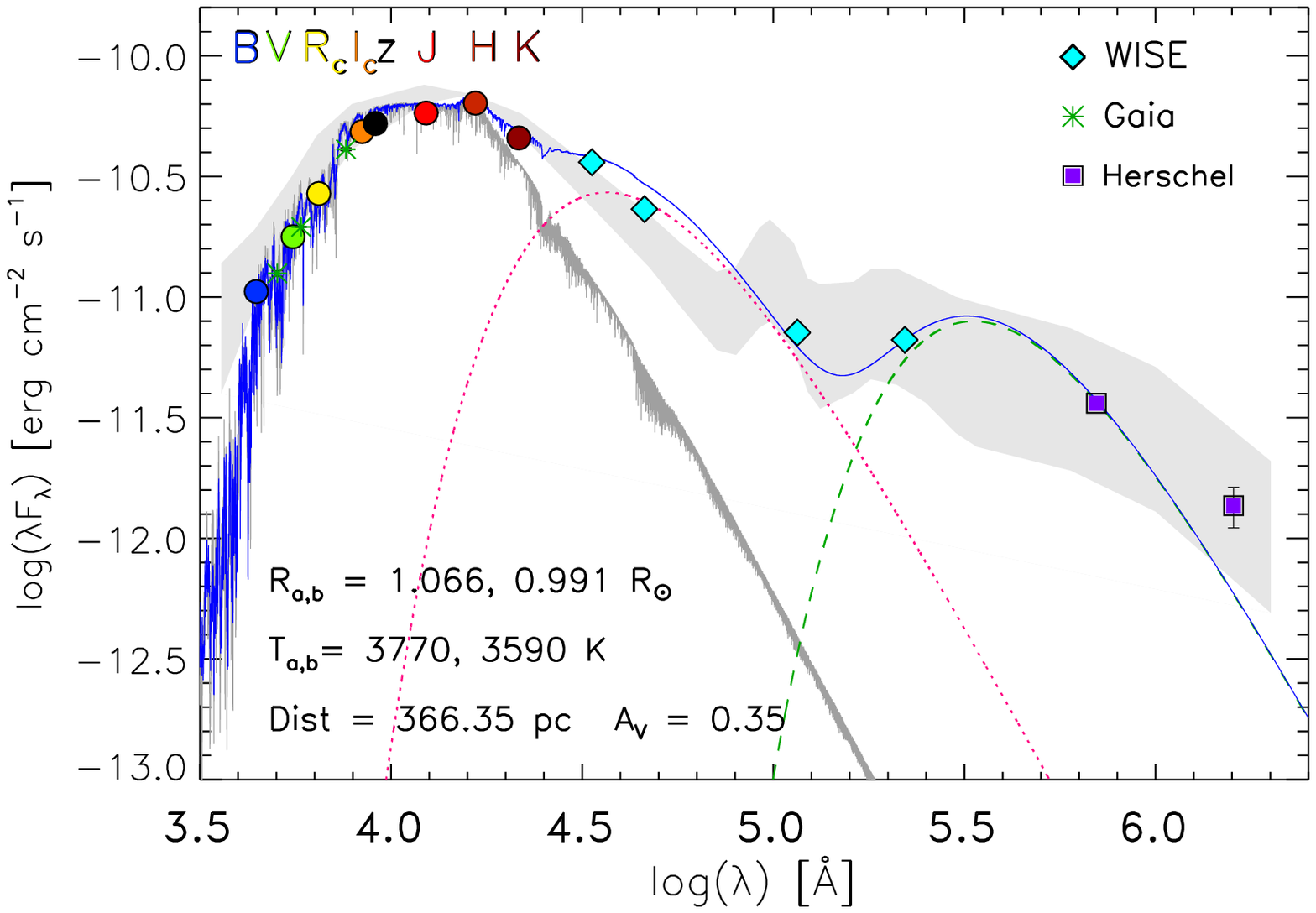}	
\caption{Spectral energy distribution of CVSO~104\,A based on OACT and synthetic photometry made on the X-shooter spectrum. {\it Gaia} magnitudes, mid- and 
far-infrared fluxes are shown  with different symbols, as indicated in the legend.
The combination of BT-Settl spectra \citep{Allard2012} that reproduce the photospheres of the components of the close binary is shown by a gray line. 
The two black bodies with $T=1000$\,K and $T=150$\,K that fit the MIR and FIR disk emission are shown by the dotted red and dashed green lines, respectively. 
The continuous blue line displays the sum of the smoothed photospheric template and the two black bodies. The light-grey shaded area in the background is the 
median SED of Class\,II Taurus sources according to \citet{DAlessio1999} and \citet{Furlan2006}, reddened by $A_V$=0.35 mag, and scaled to the $H$ band photometric point.
}
\label{fig:SED}
\end{figure}

\begin{figure}[ht]
\hspace{-0.3cm}
\includegraphics[width=9.4cm]{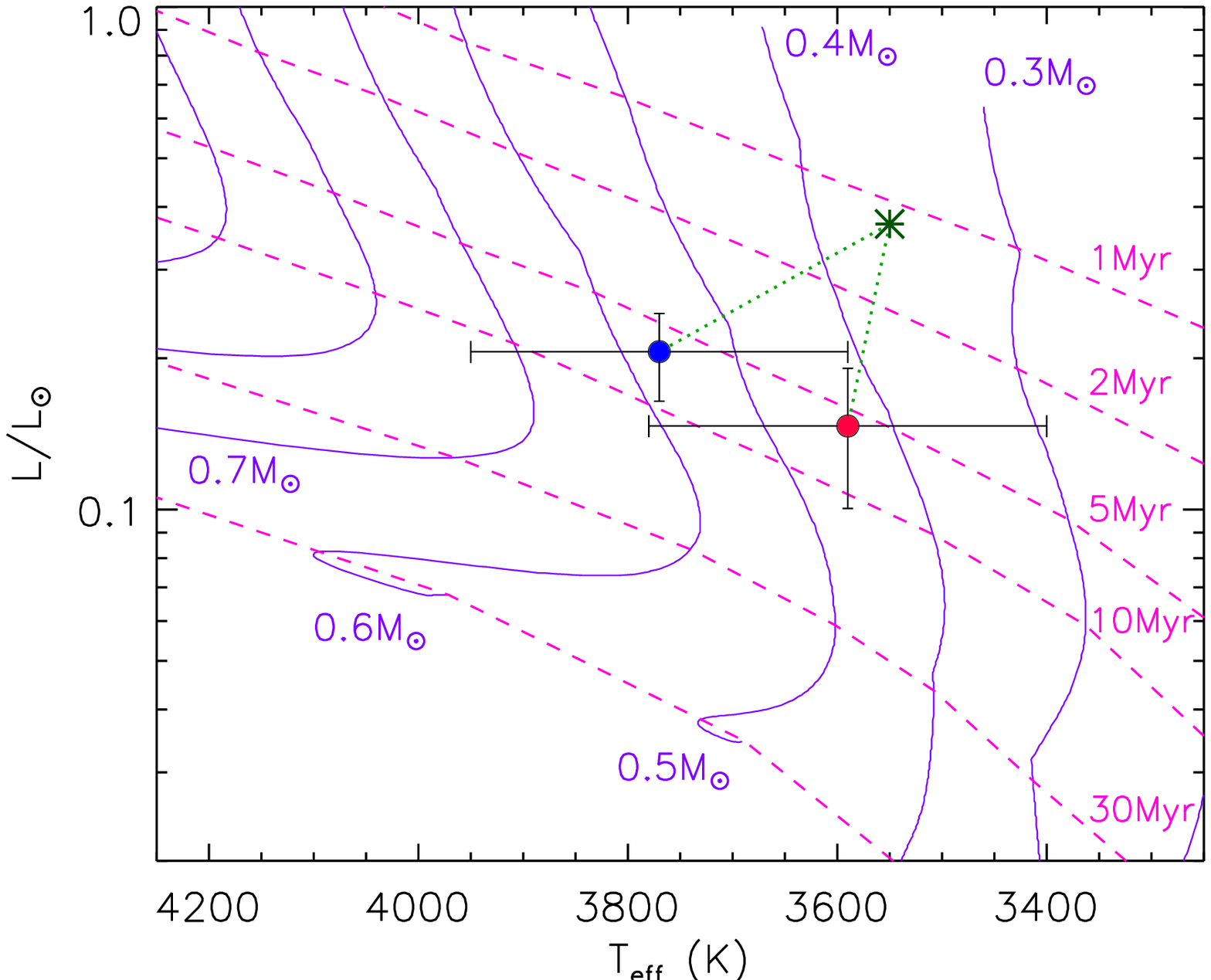}	
\caption{Position of the primary (blue dot) and secondary (red dot) component of CVSO~104\,A in the HR diagram. Isochrones and evolutionary tracks by \citet{Baraffe15} 
are overlaid as dashed and solid lines; the labels represent their age and mass.
The green asterisk connected with green dotted lines to the dots marks the position of CVSO~104\,A with the parameters derived from the combined spectrum in Paper~I.
}
\label{fig:HR}
\end{figure}

To reproduce the photospheres of the two components of the binary, we combined two BT-Settl spectra \citep{Allard2012} adopting the temperatures and flux contributions at 
red wavelengths, $w_{\rm a}$ and $w_{\rm b}$, found with \COMPO\ and reported in Table\,\ref{Tab:Param}. 
With this photospheric template, we fitted the optical-NIR portion (from $B$ to $H$ band) of the SED (Fig.~\ref{fig:SED}) fixing the  {\it Gaia}\,EDR3 parallax and 
letting the extinction $A_V$ and the radius of the primary component, $R_{\rm a}$, free to vary until a minimum $\chi^2$ is attained. We found $A_V=0.35$\,mag and 
$R_{\rm a}=1.07 R_{\sun}$.
The 3D extinction map of the Galaxy published by \citet{Green2019} provides $E(g-r)=0.12$\,mag at the position and distance of 
CVSO~104\,A\footnote{http://argonaut.skymaps.info/query}, which translates into $E(B-V)=$0.106--0.120\,mag, depending on the conversion used, which would then 
correspond to $A_V=0.33$--0.37\,mag. This value is in close agreement with the value of $A_V$ derived by us and suggests that most (or all) of this reddening is 
interstellar in nature, rather than circumstellar.
The radius of the secondary component, $R_{\rm b}=0.99 R_{\sun}$, is derived from $R_{\rm a}$ and the flux contributions $w_{\rm a}$ and $w_{\rm b}$. We have then 
evaluated the stellar luminosities as $L=4\pi R^2\sigma$\teff$^4$. The error of luminosity includes also the error on flux contribution. These stellar parameters are 
also quoted in Table\,\ref{Tab:Param}. 
The position of the two components of CVSO~104\,A in the Hertzsprung-Russell (HR) diagram is shown in Fig.~\ref{fig:HR} along with the pre-main-sequence evolutionary 
tracks and isochrones by \citet{Baraffe15}.
Both components lie close to the isochrone at 5\,Myr and masses of 0.57$\pm$0.15\,$M_{\sun}$ and 0.43$\pm$0.15\,$M_{\sun}$ can be inferred for the primary and secondary 
component, respectively. Comparing these values with the dynamical masses reported in Table\,\ref{Tab:Param}, we deduce a system inclination $i\simeq43^{+8}_{-4}$ degrees.
We note that the mass ratio derived from the position in the HR diagram,  $M_{\rm b}/M_{\rm a}=0.75\pm0.33$, is smaller than, but still compatible with the dynamical one, 
if we take the large errors of the individual masses ($\pm0.15 M_{\sun}$) into account. The latter are mainly the result of the large \teff\ uncertainty. 
Therefore, the radial velocity curve strongly suggests components with closer masses and effective temperatures.

\begin{figure}[t]
\vspace{7cm}
\includegraphics[width=9.cm]{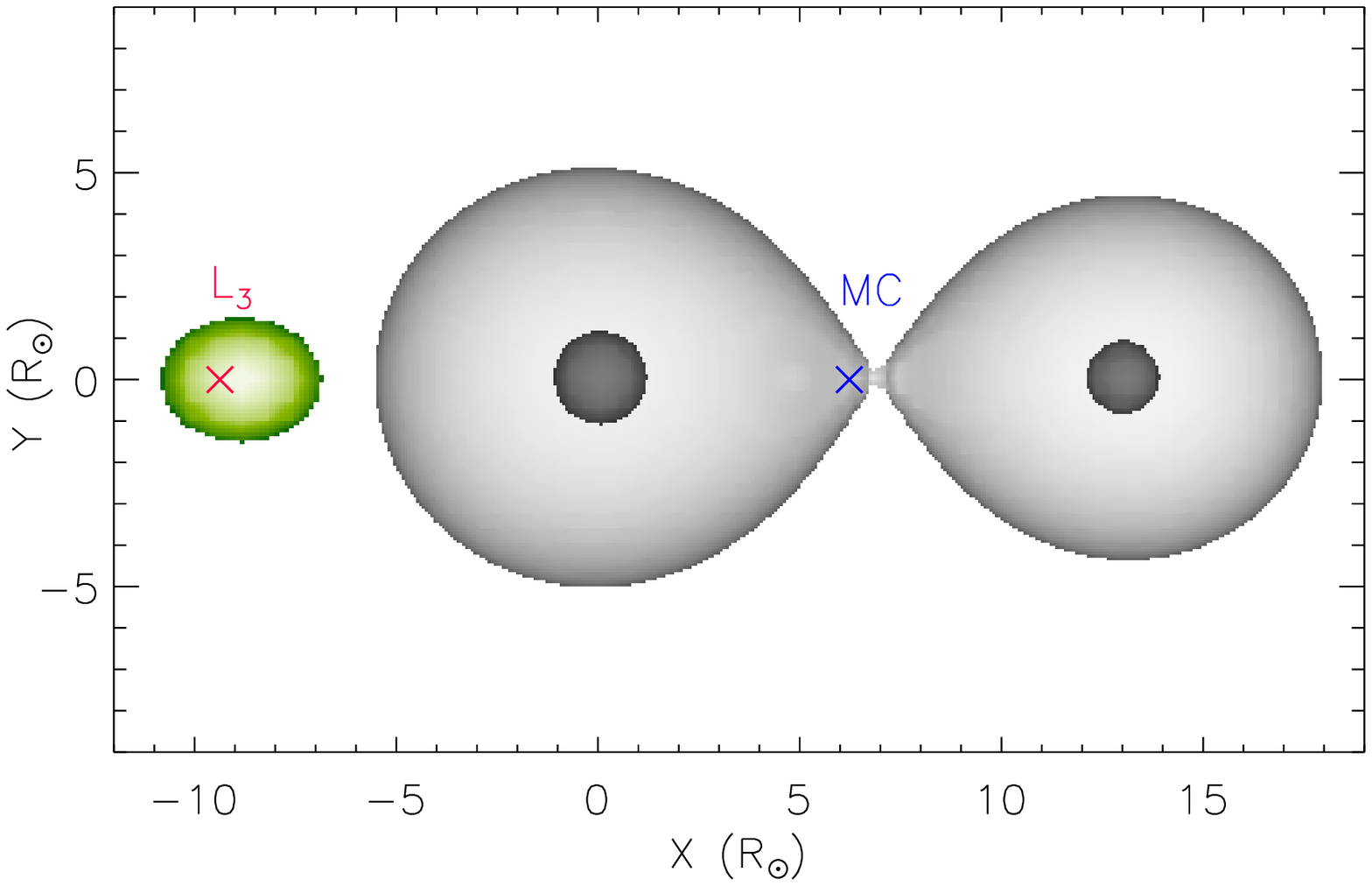}	
\caption{Schematic representation of the geometry of the CVSO~104\,A system in the orbital plane. The dark shaded surfaces represent the primary (at left) and secondary 
component, while the light-grey one is the critical Roche surface. The size and separation of the stars and Roche lobes as well as the position of the barycenter (MC) 
and the Lagrangian point $L_3$ are to scale. The source of the excess H$\alpha$ and H$\beta$ emission is represented by the shaded green area. }
\label{fig:system_conf}
\end{figure}

The masses and radii derived for the components Aa and Ab of CVSO~104 enable to trace the configuration of the system, which is depicted in Fig.~\ref{fig:system_conf}. 

The SED also shows that up to the $H$ band the emission is photospheric, while longward of that it follows the typical emission pattern displayed by Class\,II sources 
in Taurus \citep[e.g.,][]{DAlessio1999,Furlan2006}, which is outlined by the light-grey shaded area in Fig~\ref{fig:SED}. 
Such an IR excess could be explained by thermal emission from a dusty disk. We can, for example, reproduce it with two blackbodies with $T_1$=1000\,K (red dotted line 
in Fig.\,\ref{fig:SED}) and $T_2$=150\,K (green dashed line). The areas of these sources, assumed as uniformly emitting, are about 1.0$\times 10^{25}$\,cm$^2$ and 
2.1$\times 10^{28}$\,cm$^2$, respectively. 
If they are related to an inner and outer disk with inclination  $i=43$\degree\ and internal radius $R_{\rm in}=1\,R_{\sun}$, their radii should be $\sim 21$ and 
960\,$R_{\sun}$. If compared with the semi-major axis $a\simeq 13^{+1}_{-2}\,R_{\sun}$, this implies that a circumbinary disk must exist.

\section{Discussion}
\label{Sec:discussion}

Our study confirms that CVSO~104\,A is a double-lined spectroscopic binary with a dynamical mass ratio around 0.92, that is slightly lower but marginally consistent 
within the uncertainties with the estimate by \citet{Kounkel19}.
With an orbital period of 5 days, but no physical companion detected  directly, by {\it Gaia} nor by Broadening Function method, the system is a rare case of a close 
binary without a detected tertiary component. Indeed, \citet{Laos20} found that 90\% of spectroscopic binaries with orbital periods between 3 and 6 days had a tertiary 
companion. 

Observations with TESS reveal variability, with an irregular, or  at most semi-regular, pattern. The fact that TESS has large pixels, so that the component B is also 
contributing to the flux, makes difficult assessing the real amplitude of the variations and whether they really originate from the spectroscopic binary. However, 
according to our classification as a slowly-rotating, low-activity G2\,V star and the photometry displayed in Fig.~\ref{fig:ground_compB}, the brightness variations 
of the component B are negligible in comparison with CVSO~104\,A.  
The observed variations look more like small-amplitude bursts (perhaps linked to accretion) than dips, especially in the second TESS light curve, simultaneous with 
PENELLOPE. The TESS light curves are reminiscent of the sources in NGC\,2264 with ``aperiodic accretion variability'' or ``stochastic variability'' presented by 
\citet{Stauffer2016} or to the ``stochastic'' or ``burster'' type of variability detected with $K2$ in Upper Sco \citep{Cody2018}. 
With the aim of distinguishing between a ``dipper'' or ``bursting'' light curve on more quantitative grounds, we have evaluated the asymmetry of the TESS light curve 
using the metric $M$ expressed by \citet{Cody2014} in their Eq.~7. After removing the linear trend, we found $M\simeq$\,0.13. According to \citet{Cody2014}, values of 
$M<-0.25$ are typical of ``bursting'' light curves, $M>+0.25$ correspond to ``dippers'', while $-0.25<M<+0.25$ indicates symmetric light curves. As a further test, we 
calculated the third moment (skewness) as $-0.45$, which also indicates a rather symmetric light curve.
We repeated this analysis with the ASAS-SN $g'$ data, which have a precision and cadence lower than TESS, but a longer time baseline and we found $M\simeq$\,0.17 and 
skewness $-0.30$, which also suggest a symmetric light curve. Therefore, we cannot classify CVSO~104\,A either as a ``burster'' or a ``dipper''. The stochastic variability 
is probably related to fluctuating accretion.   
However, our multiband photometry (Fig.~\ref{fig:ground_LC}) shows at least two clearly detected bursts simultaneous with the enhancements of the TESS light curve. 
The intensification is stronger in the bluer bands, as usually observed during accretion bursts \citep[see, e.g.,][]{Tofflemire2017}.

In Paper\,I we quote a mass for the star of 0.37 $M_\odot$ and a mass accretion rate of $3.24\times 10^{-9} M_\odot$yr$^{-1}$, but these values assumed that the object 
was single. The mass accretion rate reported in Paper~I is higher than the \Macc\ values of each component, but it is close to their sum at RJD=59178, 
\Macc\,$\simeq2.95\times 10^{-9} M_\odot$yr$^{-1}$.

Using evolutionary tracks and isochrones of \citet{Baraffe15}, we found that both components are near the isochrone at 5\,Myr and derived the masses to be 
0.57$\pm$0.15\,$M_{\sun}$ and 0.43$\pm$0.15\,$M_{\sun}$, with a binary inclination of $i\simeq43^\circ \pm6^\circ$ (see Fig.\,\ref{fig:HR}). 
However, apart from the large mass errors, which mainly stem from the large uncertainties of \teff\ values, the masses depend on the adopted set of evolutionary tracks. 
To evaluate the impact of different models on the determination of masses and age, we have also used the SPOTS Models of \citet{Somers2020}, which include the effects 
of magnetic activity and starspots on the structure of active stars. The HR diagrams with the \citet{Somers2020} tracks for a spot covering fraction of 0\%, 34\%, and 
51\% are shown in Fig.~\ref{Fig:HR_Somers}. We note that the position of the two components with respect to the tracks with a spot filling factor $F_{\rm spot}=0$ is 
the same as for the tracks of \citet{Baraffe15}.
A spot coverage of 34\% would increase the masses by a factor of $\sim1.3$, which is inside the mass errors, while a larger spot filling factor, $F_{\rm spot}=51\%$, 
would make the masses larger by a factor $\sim1.5$ and the inclination of the system would be $i\simeq 37$\degree. However, unless the spots are evenly distributed over 
the surface of the two stars, there is no indication of such a high coverage from the light curve which should have shown a large-amplitude rotational modulation 
superimposed to the observed bursts. Moreover, in the latter case, the age of the system would turn out to be 15--20\,Myr, which is hardly compatible with the high mass 
accretion rate and with the age of $\sim$\,5\,Myr of the Ori OB1 association. 

If we use the radii and orbital period listed in Table\,\ref{Tab:Param} we find equatorial rotation velocities of  $v_{eq}=10.7$ and 10.0 \kms for the primary and 
secondary components, respectively, which become \vsini=7.3 and 6.8 \kms, i.e. in agreement with the measured \vsini, within the errors. We therefore conclude that the 
components are synchronized or close to synchronization, but the system is not yet circularized.  
This is in agreement with the time scales for circularization and synchronization, $\tau_{\rm circ}\sim 800$\,Myr and $\tau_{\rm sync}\sim 2$\,Myr, which we have 
calculated according to \citet{Zahn1989}. With an age of $\sim 5$\,Myr, this system should have had time enough to synchronize the rotation of the two components with 
the orbital period. However, as pointed out by \citet{Hut1981}, in an eccentric orbit the tidal interaction is stronger at periastron, when the orbital velocity is higher, 
with the consequence that the equilibrium is reached at a value of rotation period, $P_{\rm pseudo}$, which is smaller than $P_{\rm orb}$, leading to a 
{\it pseudo-synchronization}. The value of $P_{\rm pseudo}$ depends on the eccentricity of the system and results to be about 2.5 days for both components of CVSO~104\,A, 
which is not consistent with the \vsini\ values measured by us. The timescale for the pseudo-synchronization can be evaluated as $\tau_{\rm pseudo}\sim 12$\,Myr, 
following the guidelines of \citet{Hut1981}. This suggest that the pseudo-synchronization equilibrium has not yet been attained for the components of CVSO~104\,A.   

With a semi-major axis of about 13 R$_\odot$ and an inclination of about 43$\degr$, the system should certainly not show any eclipses. In all cases, the Roche lobe 
radius of both stars is about 5~R$_\odot$, that is, much larger than the radius of the individual stars, which is about 1~R$_\odot$ (see also Fig.~\ref{fig:system_conf}). 
There is thus also place for circumstellar accretion discs, in addition to a circumbinary disc. The existence of the latter is confirmed by our SED analysis.

The accretion luminosities and mass accretion rates of the two components, calculated from the fluxes of the lines where the profiles of the two components could be 
deblended (\ion{He}{i}$\lambda$6678, \ion{He}{i}$\lambda$5876, and H$\beta$), are similar. 
The secondary component seems to accrete slightly more than the primary, but at the first epoch (RJD= 59178) we see the reverse. However, these variations are within 
the errors and cannot be considered as highly significant.

The numerical simulations of circumbinary accretion onto eccentric and circular binaries made by \citet{Munoz2016} show that for circular binaries one expects accretion 
bursts with a period $\sim 5P_{\rm orb}$, while for eccentric orbits it is mostly modulated at $\sim 1P_{\rm orb}$. This seems to be the case with CVSO~104\,A, according 
to what is observed in the TESS light curve (Fig.~\ref{fig:TESS}). Moreover, based on the above simulations, the two components should have similar accretion rates in 
circular orbits, while very different accretion rates, with a ratio up to 10--20, should be observed in the components of binaries with eccentric orbits. 
The asymmetry breaking between the stars, however, alternates over timescales of the order of 200\,$P_{\rm orb}$ and it can be attributed to a slowly precessing, eccentric 
circumbinary disk. Our spectroscopic observations always display very similar \Macc\ for the two components, with some hint of variation, but they  span a too small time 
range and further observations are needed to search for variations on 100-day timescales.  

Figure\,1 of \citet{Munoz2016}  clearly shows the spiral pattern of matter from the circumbinary disk to the stars/circumstellar disks. 
Similar structures are found in the simulations of \citet{Gillen2017} and \citet{Val-Borro2011} for the two gas streams passing the co-linear Lagrangian points on the 
way down to the stars. \citet{Val-Borro2011} also used their model to estimate the RV of Balmer line emission from the region within the circumbinary disk, finding maximum 
velocities of about $\pm100$\,\kms\ around the quadratures, in agreement with the excess emission observed on V4046\,Sgr by \citet{Stempels2004}. According to 
\citet{Stempels2004}, these excess emissions are consistent with two concentrations of gas co-rotating with the stars and moving with a projected velocity of 80\,\kms\ 
around the center of mass. This suggests that they are located well inside the edge of the circumbinary disk, and also inside the co-linear Lagrangian points of V4046\,Sgr. 
It is possible that a similar accretion structure that brings matter towards the primary component is responsible for the extra emission component observed in H$\alpha$ 
and H$\beta$. We note that its RV is approximately symmetric with respect to barycentric velocity $\gamma=24.5$\,\kms. If we are looking at the same structure at opposite 
phases (near quadratures) and if we assume it as quasi-stationary in the reference frame rotating with the system, it should be located at $\approx 15 R_{\sun}$ from the 
barycenter, i.e. near the Lagrangian point $L_3$ (see Fig.~\ref{fig:system_conf}).  
The absence of a similar structure on the side of the secondary star is not an unexpected result, since the simulations of \citet{Val-Borro2011} and \citet{Munoz2016} show 
that the density distribution in the inner gap can be highly asymmetrical. For instance, the model of \citet{Terquem2015} applied to CoRoT 223992193, a PMS binary with 
$P_{\rm orb}\simeq 3.87$\,day and M-type components, shows that the stream of matter from the circumbinary disk to the primary is much denser than the one directed toward 
the secondary component. 
A similar result is also reported by \citet{Gomez2020} for simulations of AK~Sco, a short-period SB2 composed of mid-F stars with a circumbinary disk. They observed an 
enhancement of the accretion rate during the periastron passages from UV tracers and higher resolution COS spectra revealed that the flow was channeled preferentially into 
one of the two components.

Contrary to what observed by \citet{Gomez2020} for AK\,Sco,  \citet{Ardila2015} do not detect any clear correlation between accretion luminosity and phase for the two 
binaries DQ~Tau and UZ~Tau either from UV continuum or \ion{C}{iv} line flux, suggesting that gas is stored in the system throughout the orbit and may accrete stochastically.  
These systems are both composed of early-M  type stars in eccentric orbits with longer orbital periods ($P_{\rm orb}=15.8$ and 19.1 day, respectively) and circumbinary disks. 
However, optical emission lines and continuum veiling intensification near the periastron passage have been reported for DQ~Tau \citep[e.g.,][]{Mathieu1997,Basri1997}. 
\citet{Kospal2018} analyzed contemporaneous ground-based, {\it Spitzer}, and $K2$ photometry of DQ~Tau. The $K2$ light curve displays a clear rotational modulation which is 
crossed by short-duration flare-like events but also by stronger brightening events with a complex behaviour and a longer duration, which are observed only at phases close 
to the periastron passage. The latter are interpreted as accretion bursts. \citet{Tofflemire2017} observed accretion bursts near periastron passages in the same binary system 
with multiband photometry which are stronger in the bluer bands. We observed a similar color behaviour for the bursts observed in CVSO~104\,A, but they are not seen near 
periastron passages. 
Another system for which accretion burst have been observed near periastron passages is TWA\,3A \citep{Tofflemire2019}. This is apparent from their $U$-band light curve, but 
it is also indicated by the intensity and FWHM of H$\alpha$, H$\beta$ and \ion{He}{i}\,$\lambda$5876 lines, which increase near the periastron passages. They also show that 
the \ion{He}{i}\,$\lambda$5876 line indicates that the primary is accreting more than the secondary and suggest that this can be explained by the \citet{Munoz2016} simulations.
We note, however, that $P_{\rm orb}$ and eccentricity are larger for TWA\,3A  than for CVSO~104\,A, with periastron and apastron separations of 14.7 and 64.0 $R_{\sun}$, 
respectively. Therefore the effect of periastron passage may be stronger for the former system.
We need time-series high-resolution spectroscopy, possibly in different epochs, to investigate the behavior of accretion variability in CVSO~104\,A.

\section{Conclusions}
\label{sect:Conclusions}

We presented a spectroscopic and photometric study of the PMS object CVSO~104\,A, which is located in the Ori OB1 association and has an optical companion  with a similar 
brightness at about 2$\farcs4$. The latter star, which we have named CVSO~104\,B, is a background Sun-like star not physically associated with the PMS object and does not 
belong to Ori OB1. Thanks to high- and intermediate-resolution spectra taken in the framework of the PENELLOPE large program and archival APOGEE spectra, we confirmed 
CVSO~104\,A as a double-lined spectroscopic binary and derived, for the first time, its orbital parameters. We found a dynamical mass ratio of 0.92, an orbital period of 
about 5 days, and an eccentric orbit ($e\simeq0.39$, see Table\,\ref{Tab:Param}). It is a rare case of a close binary without a detected tertiary component.

The analysis of the UVES spectra allowed also us to estimate the parameters of the two components of the binary system, which turn out to be slowly-rotating 
(\vsini\,$\simeq6-7$\,\kms), early M-type stars. In the   HR diagram both stars are located near the same isochrone at 5~Myr (according to two different models), in good 
agreement with the age estimated for Ori OB1. The rotation rates indicate that the two stars have already attained  spin-orbit synchronization but the orbit is not yet 
circular. This agrees with the timescales for synchronization and circularization that we have estimated as $\tau_{\rm sync}\sim 2$\,Myr and  $\tau_{\rm circ}\sim 800$\,Myr, 
respectively.

The SED shows a significant infrared excess that can be explained only with a circumbinary accretion disk with an extension of at least 4.5 au (960 $R_{\sun}$), although 
the presence of circumstellar disks around the two stars cannot be ruled out. The kinematic properties (RV and FWHM) of the narrow component of the 
[\ion{O}{i}]\,$\lambda$6300\AA\ line are also compatible with emission from a region of a circumbinary disk of $\approx$ 1 au.

The analysis of permitted lines, such as H$\beta$, \ion{He}{i}\,$\lambda$5876\,\AA\, and \ion{He}{i}\,$\lambda$6678\,\AA, after the removal of the underlying photospheric 
spectrum, clearly reveals emission from both components of the close binary CVSO~104\,A with a similar intensity. This result is in contrast with what is suggested by some 
theoretical studies, which predict, for most of the time, very different accretion rates for the components of an eccentric binary system, even if they have a similar mass 
\citep[e.g.,][]{Munoz2016}. However, since the ratio of accretion rates of the components is expected to reverse periodically, a binary system should be observed 
nearly-continuously for a long time ($\gtrsim 100$\,d) to draw firm conclusions. 
In addition to the emission profiles corresponding to the velocity of the components and likely produced by emitting material near the accretion shocks, the H$\alpha$ and 
H$\beta$ profiles display a broad excess emission component, which appears blueshifted or redshifted by more than 100\,\kms\ at different phases. We think that this emission 
could be produced by 
an accretion funnel from the circumbinary disk towards the primary component, similar to what found by \citet{Stempels2004} for V4046~Sgr and in agreement with the predictions 
by numerical simulations \citep[e.g.,][]{Val-Borro2011,Terquem2015,Gillen2017}.

The contemporaneous space- (TESS) and ground-based  photometry displays a stochastic variability pattern with a possible periodicity around 4.7 days (not far from the orbital 
period) and some short-duration ($\approx$ 1 day) peaks. The latter are reminiscent of accretion bursts, based on their shape and color dependence. However, unlike other 
binary systems with eccentric orbits (such as the paradigmatic case of DQ~Tau), these accretion bursts do not seem to occur near the periastron passages. 

Future studies with multi-epochs and high-resolution spectroscopy of this and other PMS binaries are needed for a deeper investigation of the impact of multiplicity on 
the mass accretion phenomenon. 
The synergy between ODYSSEUS and PENELLOPE large programs can be of great help in this respect and the data are publicly 
available\footnote{\tt https://zenodo.org/communities/odysseus/?page=1\&size=20}.

\begin{acknowledgements}
We thank the anonymous referee for her/his useful comments and suggestions.
We acknowledge the support from the Italian {\it Ministero dell'Istruzione, Universit\`a e  Ricerca} (MIUR).
This work has been partially  supported by the project  PRIN-INAF-MAIN-STREAM 2017
``Protoplanetary disks seen through the eyes of new-generation instruments'' and from the project PRIN-INAF 2019
 "Spectroscopically Tracing the Disk Dispersal Evolution".
This work benefited from discussions with the ODYSSEUS team (HST AR-16129), {\tt https://sites.bu.edu/odysseus/}.
This project has received funding from the European Union's Horizon 2020 research and innovation programme under the Marie Sk\l odowska-Curie grant agreement 
No 823823 (DUSTBUSTERS). This project has received funding from the European Research Council (ERC) under the European Union's Horizon 2020 research and innovation 
programme under grant agreement No 716155 (SACCRED). This work was partly supported by the Deutsche Forschungs-Gemeinschaft (DFG, German Research Foundation) - 
Ref no. FOR 2634/1 TE 1024/1-1. FMW is grateful to the AAVSO for the award of AAVSOnet telescope time, and to the Ken Menzies for implementing the observations.
We thank Elizabeth Waagen and the citizen scientists of the AAVSO for their contributions to this program.
K.G. acknowledges the partial support from the Ministry of Science
and Higher Education of the Russian Federation (grant 075-15-2020-780).
This research made use of SIMBAD and VIZIER databases, operated at the CDS, Strasbourg, France, and of {\sc Lightkurve}, a Python package for Kepler and TESS data 
analysis \citep{2018ascl.soft12013L}. 

\end{acknowledgements}

\bibliography{bibliography.bib}

\newpage
\begin{appendix}

\section{Additional tables and figures}
\label{sec:appendix}

\begin{figure}
\begin{center}
\hspace{-.7cm}
\includegraphics[width=9.3cm]{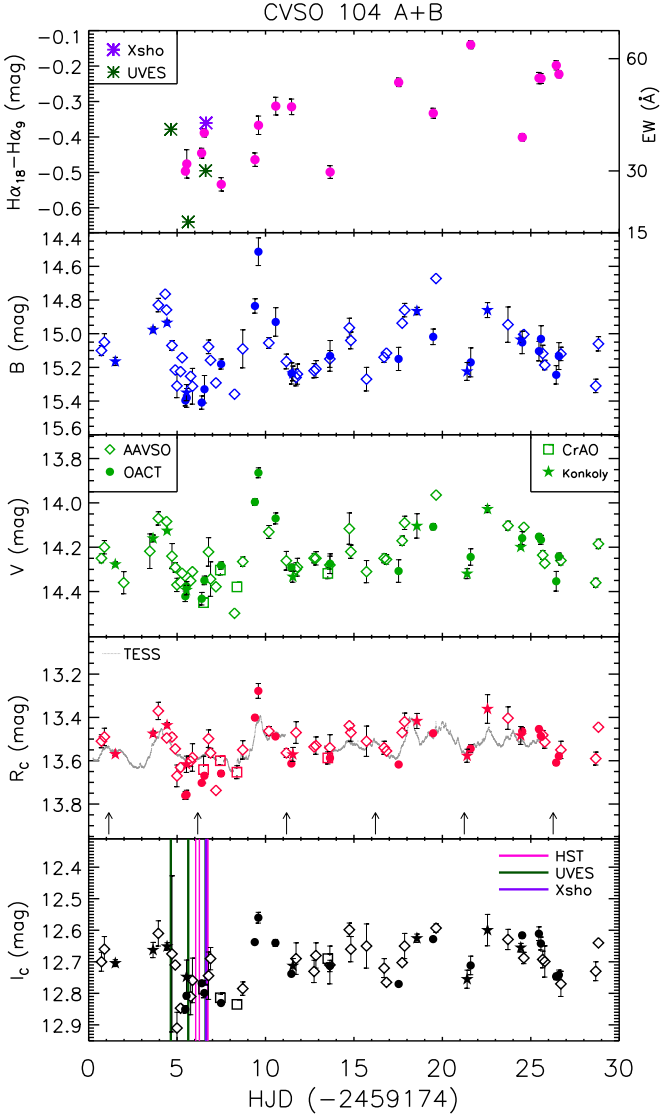}	
\caption{Multiband ground-based light curve of CVSO~104\,A+B around the time of VLT and HST observations. Different symbols, as reported in the legend, are used 
for the different data sets. The H$\alpha$ color index, which measures the intensity of the line, is shown in the upper panel. Synthetic H$\alpha$ colors based on 
X-Shooter and UVES spectra are overplotted with asterisks in the same box.
The contemporaneous TESS light curve is overplotted with grey dots to the $R_{\rm C}$ light curve. The arrows in the same box mark the periastron passages. 
The epoch of VLT and HST observations are marked in the lower box.  
}
\label{fig:ground_LC}
\end{center}
\end{figure}

\begin{figure}
\begin{center}
\hspace{-.7cm}
\includegraphics[width=9.3cm]{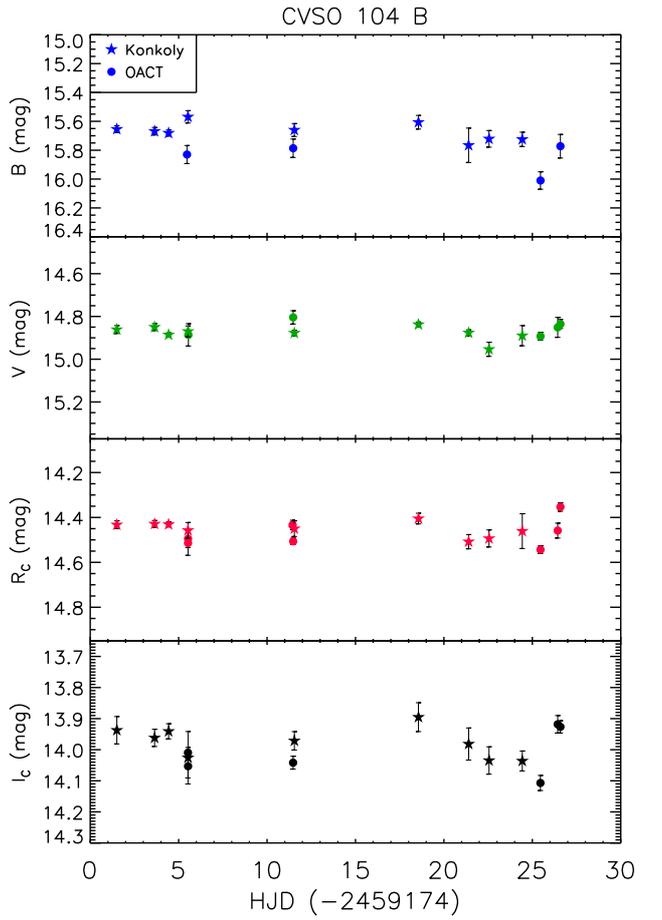}	
\caption{Multiband ground-based photometry of CVSO~104\,B obtained at OACT and Konkoly observatories in the nights with a good seeing.  
}
\label{fig:ground_compB}
\end{center}
\end{figure}

\setlength{\tabcolsep}{3pt}

\begin{table}
\caption{Photometry of CVSO~104\,A and B. }	
\begin{tabular}{lcccl}   
\hline\hline
\noalign{\smallskip}
Band  & $\lambda_{\rm c}$ &  Comp.\,A & Comp.\,B  &   Reference   \\             
      & ($\mu$m) & (mag) & (mag) &  \\ 
\hline
\noalign{\smallskip}
 $B$      &  0.444  & 16.67$\pm$0.08  & 15.67$\pm$0.06 & \scriptsize{Present work} \\ 
 $V$      &  0.550  & 15.57$\pm$0.05  & 14.87$\pm$0.04 & \scriptsize{Present work}   \\
 $R_{C}$  &  0.621  & 14.64$\pm$0.05  & 14.47$\pm$0.05 & \scriptsize{Present work}   \\
 $I_{C}$  &  0.767  & 13.53$\pm$0.04  & 14.10$\pm$0.07 & \scriptsize{Present work}   \\
 $z'$     &  0.910  & 13.56$\pm$0.05  & 14.43$\pm$0.11 & \scriptsize{Present work}   \\
 $BP$     &  0.505  & 15.242$\pm$0.042 & 15.024$\pm$0.004 & {\it Gaia}\,EDR3 \\ 
 $G$      &  0.623  & 14.455$\pm$0.009 & 14.588$\pm$0.003 & {\it Gaia}\,EDR3 \\ 
 $RP$     &  0.772  & 13.398$\pm$0.018 & 13.932$\pm$0.005 & {\it Gaia}\,EDR3 \\ 
 $J$      &  1.235   & 12.16$\pm$0.05  & 13.52$\pm$0.08  & \scriptsize{Present work} \\
 $H$      &  1.662   & 11.24$\pm$0.05  & 13.03$\pm$0.08  & \scriptsize{Present work}   \\
 $K_{\rm s}$     &  2.159   & 10.81$\pm$0.04  & 13.23$\pm$0.08  & \scriptsize{Present work}   \\
$WISE\,1$ &  3.35   & ~9.699$\pm$0.023 & \dots &  WISE \\ 
$WISE\,2$ &  4.60   & ~9.204$\pm$0.020 & \dots &  WISE \\ 
$WISE\,3$ & 11.56   & ~7.560$\pm$0.019 & \dots  &  WISE \\ 
$WISE\,4$ & 22.09   & ~5.571$\pm$0.039 & \dots  &  WISE \\ 
$Herschel\,blue$ &  70 & 84.7$\pm$2.7 mJy  & \dots &   H2020 \\ 
$Herschel\,red$ &  160 & 72.9$\pm$14.0 mJy & \dots &   H2020 \\ 
 \noalign{\smallskip}
 \hline  
\normalsize
\end{tabular}
~\\ \textbf{Notes}  {\it Gaia}\,EDR3\,=\,\citet{GaiaDR3}; WISE\,=\,\citet{WISE};
H2020\,=\,\citet{2020yCat.8106....0H}.
\label{Tab:SED}
\end{table}

\begin{figure}
\begin{center}
\hspace{-.7cm}
\includegraphics[width=9.3cm]{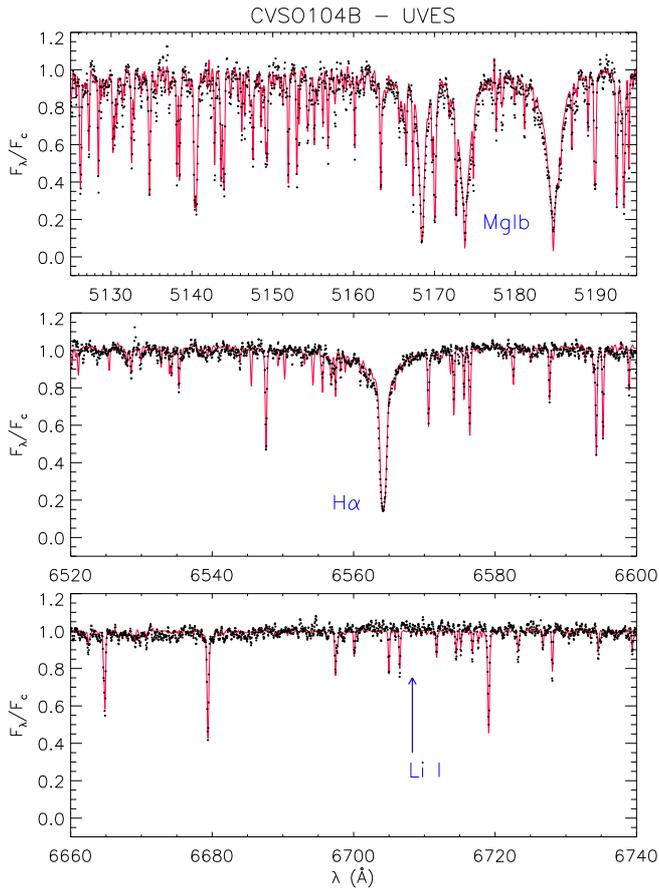}	
\caption{UVES spectrum of CVSO~104\,B (black dots) in three spectral regions around the \ion{Mg}{i}\,b triplet ({\it top}), H$\alpha$ ({\it middle}), and 6700 \AA\ 
({\it bottom}). In each box the spectrum of the standard star that is best fitting that of CVSO~104\,B is overplotted with a full red line. 
}
\label{Fig:spe_B}
\end{center}
\end{figure}

\begin{figure}
\begin{center}
\hspace{-0.5cm}
\includegraphics[width=9.cm]{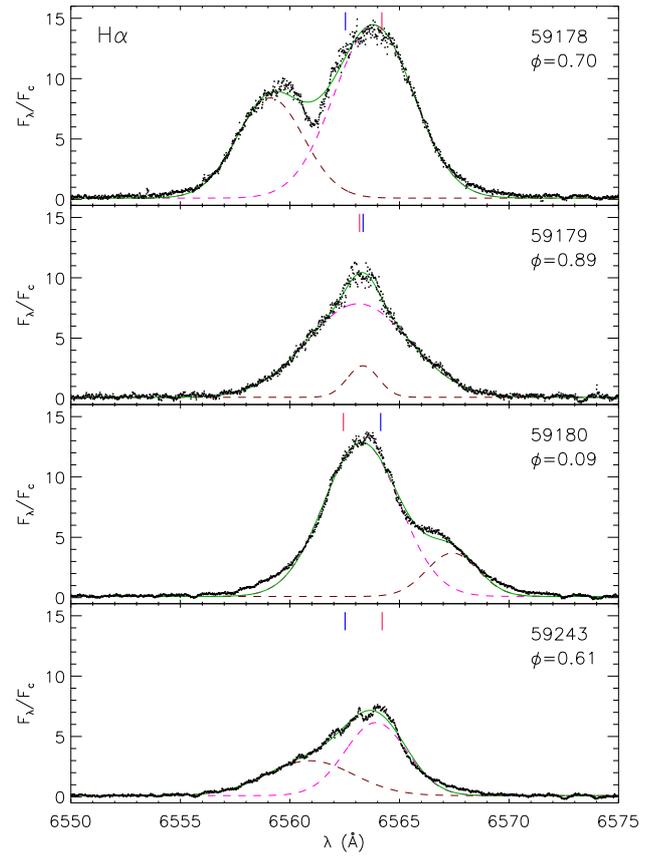}	
\vspace{0cm}
\caption{Residual UVES H$\alpha$ profiles (black dots). The totally  blended emission from the two stars is represented, in each box, by a magenta dashed line, 
while the brown dashed line is the Gaussian fitted to the excess blueshifted or redshifted emission. The sum of the two Gaussians is overplotted as a full green line. 
The blue and red vertical ticks mark the expected position of the lines of primary and secondary component, respectively, according to the photospheric RV. }
\label{fig:Halpha}
\end{center}
\end{figure}

\begin{figure}[ht]
\includegraphics[width=8cm]{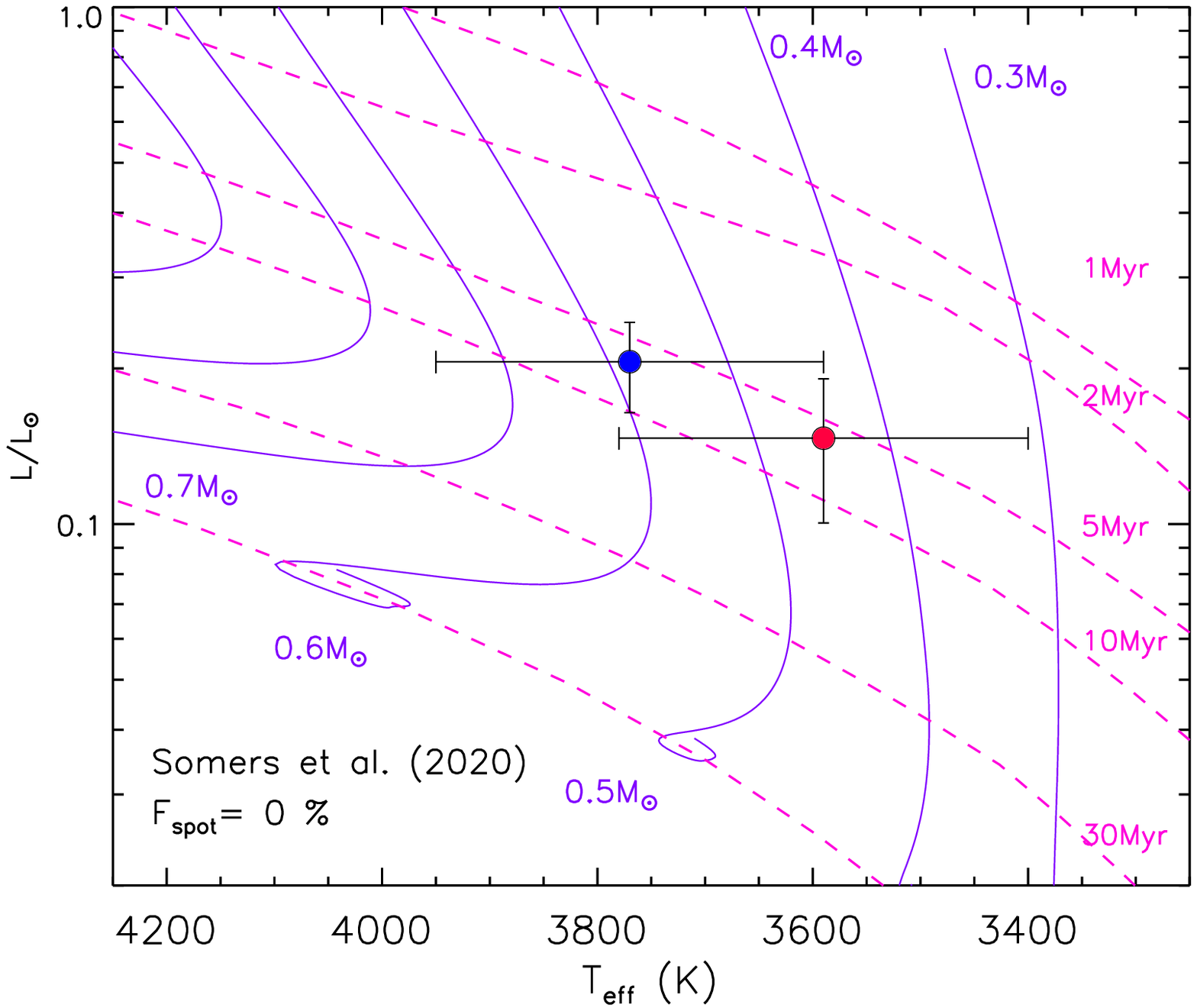}	
\includegraphics[width=8cm]{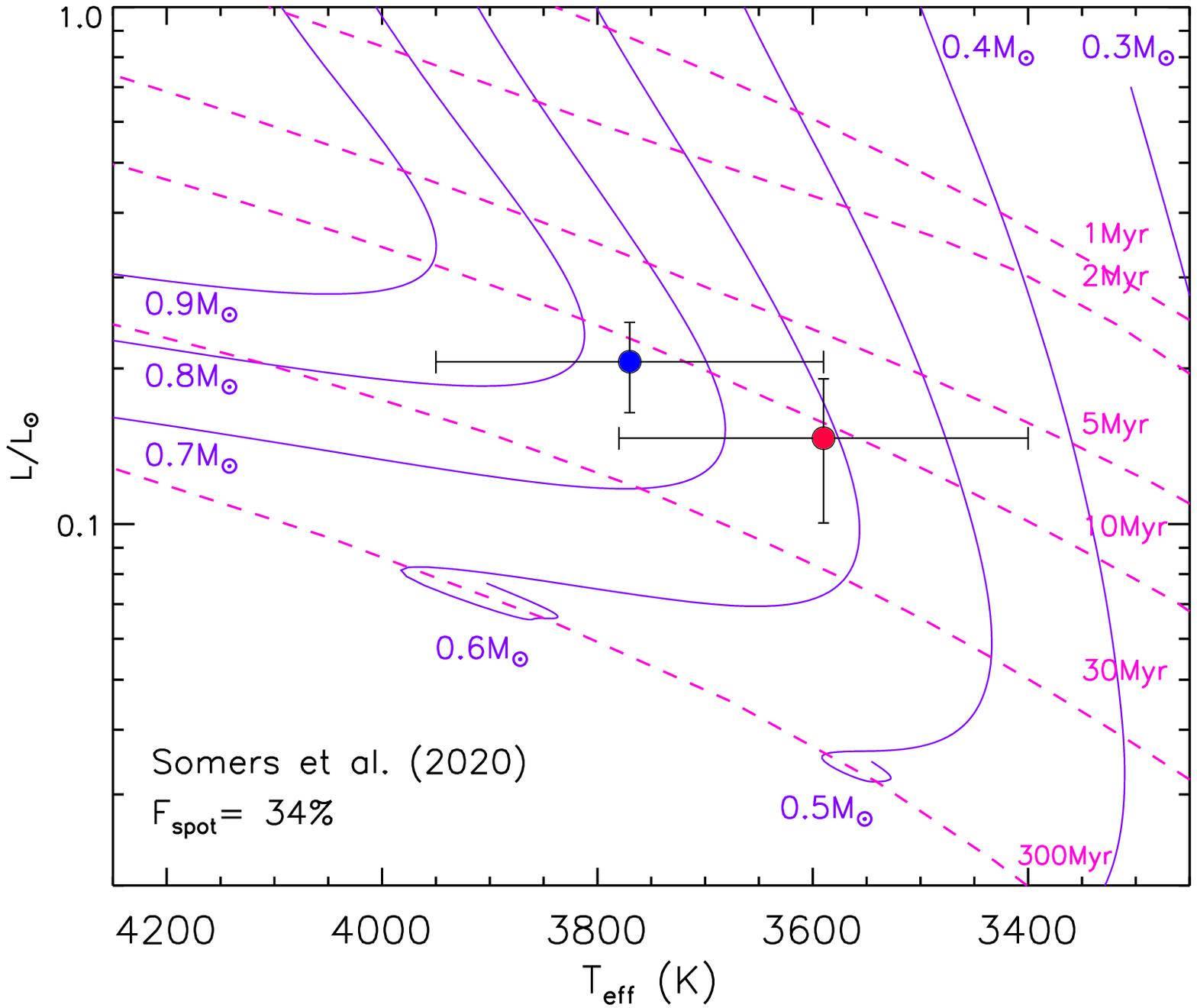}	
\includegraphics[width=8cm]{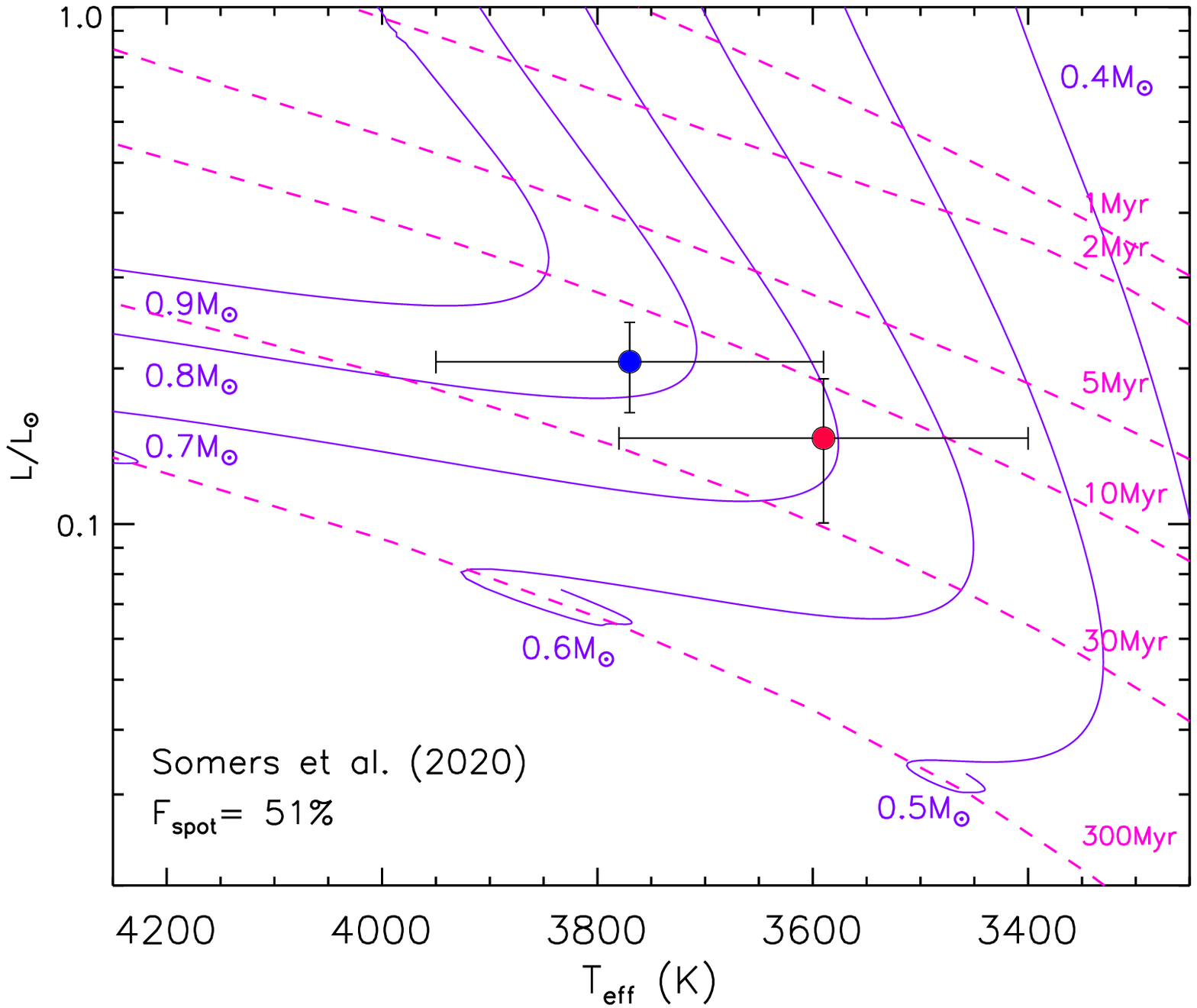}	
\caption{HR diagram of the primary (blue dot) and secondary (red dot) component of CVSO~104\,A  with isochrones and evolutionary tracks by \citet{Somers2020} for 
three spot covering factors.
}
\label{Fig:HR_Somers}
\end{figure}

\end{appendix}

\end{document}